\newif\ifthesis
\thesisfalse
\newif\ifwiley
\wileyfalse

\documentclass[10pt,journal,twocolumn]{IEEEtran}

\usepackage{wrapfig,booktabs,amsmath,amsthm,amssymb}
\ifwiley\else\usepackage{cite}\fi
\usepackage{bm,url,multirow,times,enumitem,comment}
\usepackage{mathtools,siunitx,balance,tikz,microtype,xcolor,xspace}
\ifwiley\else\usepackage{mathabx}\fi
\usepackage[font=footnotesize]{subcaption}
\usepackage{import}
\usepackage{graphicx} 
\usepackage{siunitx}
\usepackage{makecell}
\usepackage{xfrac}
\usepackage{pdflscape}
\usepackage{fancyhdr}

\ifwiley
	\newcommand\Earth{\oplus}
\fi

	\usepackage[linesnumbered,ruled]{algorithm2e}

\ifwiley
\else
	\newcommand{\citep}{\cite}
\fi

\ifthesis
	\usepackage[colorlinks=true, linkcolor=black, citecolor=black, urlcolor=blue]{hyperref}
\else
	\ifwiley
	\else
		\usepackage[colorlinks=true, linkcolor=black, citecolor=blue, urlcolor=blue]{hyperref}
	\fi
\fi

\ifthesis
	\newcommand{\paperOrThesis}{thesis}
\else
	\newcommand{\paperOrThesis}{paper}
\fi

\newcommand{\vb}{\boldsymbol}
\newcommand{\vbh}[1]{\hat{\boldsymbol{#1}}}
\newcommand{\vbc}[1]{\check{\boldsymbol{#1}}}
\newcommand{\vbb}[1]{\bar{\boldsymbol{#1}}}
\newcommand{\vbt}[1]{\tilde{\boldsymbol{#1}}}
\newcommand{\vbs}[1]{{\boldsymbol{#1}}^*}

\newcommand{\tr}{\mathsf{T}}
\newcommand{\ba}{\begin{array}}
\newcommand{\ea}{\end{array}}

\newcommand{\define}{\overset{\Delta}{=}}
\renewcommand{\equiv}{\define}

\newcommand{\E}[1]{\mathbb{E}\left[ #1 \right]}

\DeclareMathAlphabet{\mathpzc}{OT1}{pzc}{m}{it}

\renewcommand{\c}{\mathrm{c}}
\newcommand{\s}{\mathrm{s}}
\newcommand{\Exp}{\mathrm{Exp}}
\newcommand{\Log}{\mathrm{Log}}
\newcommand{\cpe}[1]{\begin{bmatrix}{#1}\times\end{bmatrix}}
\newcommand{\mr}[1]{{\mathrm{#1}}}
\newcommand{\norm}[1]{\left\lVert#1\right\rVert}
\newcommand{\snorm}[1]{\lVert{#1}\rVert}
\newcommand{\chol}[1]{\texttt{chol}\left({#1}\right)}
\newcommand{\pdv}[2]{\frac{\partial{#1}}{\partial{#2}}}

\newcommand{\pprx}{PpRx\xspace}
\newcommand{\ppe}{PpEngine\xspace}

\ExplSyntaxOn 
\cs_new:Npn \expandableinput #1
  { \use:c { @@input } { \file_full_name:n {#1} } }
\ExplSyntaxOff

\ifthesis
	\newenvironment{inlinealgo}[1]{
		\begin{algorithm}
			\SetKwInOut{Input}{Input}
			\SetKwInOut{Output}{Output}	
		\caption{\texttt{#1}}
	}{\end{algorithm}}
\else
		\newenvironment{inlinealgo}[1]%
		{
			\refstepcounter{algocf}
			\vspace{10pt}
			\SetKwInOut{Input}{Input} 
			\SetKwInOut{Output}{Output}
			\SetAlgoNoLine 
			\hrule height 0.75pt
			\vspace{2pt}%
			\textbf{Algorithm {\thealgocf}:} \texttt{#1}%
			\vspace{2pt}%
			\hrule height 0.75pt
			\vspace{2pt}%
		}
		{{\vspace{5pt}}%
			\hrule height 0.75pt%
			\vspace{10pt}%
		}
\fi

\renewcommand{\frame}[1]{\boldsymbol{\mathrm{#1}}}
\newcommand{\vecframestyle}[1]{\boldsymbol{\mathrm{#1}}}

\newcommand{\vecf}[4]{\vb{#1}^{\vecframestyle{#2}\vecframestyle{#3}\vecframestyle{#4}}}

\newcommand{\vecfh}[4]{\vbh{#1}^{\vecframestyle{#2}\vecframestyle{#3}\vecframestyle{#4}}}

\renewcommand{\vec}[2]{\vecf{#1}{#2}{}{}}

\newcommand{\vech}[2]{\vecfh{#1}{#2}{}{}}

\newcommand{\vecs}[3]{\vb{#1}_{#3}^{\vecframestyle{#2}}}

\newcommand{\rots}[4]{\vb{#1}_{#4}^{\vecframestyle{#2}\vecframestyle{#3}}}
\newcommand{\rot}[3]{\rots{#1}{#2}{#3}{}}
\newcommand{\R}[2]{\rot{R}{#1}{#2}}
\newcommand{\Rs}[3]{\rots{R}{#1}{#2}{#3}}

\newcommand{\baselinemayninelordPv}{97.28\%}

\newcommand{\baselinemayninelordPf}{0.21\%}

\newcommand{\baselinemayninelordpnftotal}{13.8}
\newcommand{\baselinemayninelordrmse}{20.6}

\newcommand{\baselinemayninelordpnfHorzTotal}{10.6}
\newcommand{\baselinemayninelordrmseHorz}{18.9}

\newcommand{\baselinemayninelordpnfVertTotal}{4.0}
\newcommand{\baselinemayninelordrmseVert}{8.1}
\newcommand{\baselinemayninelordrmsPitch}{0.18}
\newcommand{\baselinemayninelordrmsRoll}{0.21}
\newcommand{\baselinemayninelordrmsYaw}{0.22}
\newcommand{\baselinemayninelordpnfPitch}{0.28}
\newcommand{\baselinemayninelordpnfRoll}{0.28}
\newcommand{\baselinemayninelordpnfYaw}{0.28}

\newcommand{\baselinemayninelynxPv}{94.56\%}

\newcommand{\baselinemayninelynxPf}{0.19\%}

\newcommand{\baselinemayninelynxpnftotal}{20.2}
\newcommand{\baselinemayninelynxrmse}{34.4}

\newcommand{\baselinemayninelynxpnfHorzTotal}{16.4}
\newcommand{\baselinemayninelynxrmseHorz}{28.6}

\newcommand{\baselinemayninelynxpnfVertTotal}{7.3}
\newcommand{\baselinemayninelynxrmseVert}{19.2}
\newcommand{\baselinemayninelynxrmsPitch}{0.60}
\newcommand{\baselinemayninelynxrmsRoll}{0.59}
\newcommand{\baselinemayninelynxrmsYaw}{0.26}
\newcommand{\baselinemayninelynxpnfPitch}{0.99}
\newcommand{\baselinemayninelynxpnfRoll}{0.99}
\newcommand{\baselinemayninelynxpnfYaw}{0.99}

\newcommand{\baselinemaytwelvelordPv}{97.74\%}

\newcommand{\baselinemaytwelvelordPf}{0.43\%}

\newcommand{\baselinemaytwelvelordpnftotal}{12.3}
\newcommand{\baselinemaytwelvelordrmse}{13.8}

\newcommand{\baselinemaytwelvelordpnfHorzTotal}{9.7}
\newcommand{\baselinemaytwelvelordrmseHorz}{12.5}

\newcommand{\baselinemaytwelvelordpnfVertTotal}{4.9}
\newcommand{\baselinemaytwelvelordrmseVert}{5.8}
\newcommand{\baselinemaytwelvelordrmsPitch}{0.17}
\newcommand{\baselinemaytwelvelordrmsRoll}{0.21}
\newcommand{\baselinemaytwelvelordrmsYaw}{0.18}
\newcommand{\baselinemaytwelvelordpnfPitch}{0.29}
\newcommand{\baselinemaytwelvelordpnfRoll}{0.29}
\newcommand{\baselinemaytwelvelordpnfYaw}{0.29}

\newcommand{\baselinemaytwelvelynxPv}{98.63\%}

\newcommand{\baselinemaytwelvelynxPf}{0.56\%}

\newcommand{\baselinemaytwelvelynxpnftotal}{13.7}
\newcommand{\baselinemaytwelvelynxrmse}{7.7}

\newcommand{\baselinemaytwelvelynxpnfHorzTotal}{10.6}
\newcommand{\baselinemaytwelvelynxrmseHorz}{6.0}

\newcommand{\baselinemaytwelvelynxpnfVertTotal}{5.7}
\newcommand{\baselinemaytwelvelynxrmseVert}{4.8}
\newcommand{\baselinemaytwelvelynxrmsPitch}{0.41}
\newcommand{\baselinemaytwelvelynxrmsRoll}{0.43}
\newcommand{\baselinemaytwelvelynxrmsYaw}{0.27}
\newcommand{\baselinemaytwelvelynxpnfPitch}{0.84}
\newcommand{\baselinemaytwelvelynxpnfRoll}{0.84}
\newcommand{\baselinemaytwelvelynxpnfYaw}{0.84}

\newcommand{\baselinelordPv}{97.52\%}

\newcommand{\baselinelordPvShort}{97.5\%}

\newcommand{\baselinelordPfShort}{0.3\%}

\newcommand{\baselinelordpnfHorzFix}{8.4}
\newcommand{\baselinelordpnfHorzTotal}{10.1}

\newcommand{\baselinelynxPv}{96.62\%}

\newcommand{\baselinelynxPvShort}{96.6\%}

\newcommand{\baselinelynxPfShort}{0.4\%}

\newcommand{\baselinelynxpnfHorzFix}{9.2}
\newcommand{\baselinelynxpnfHorzTotal}{12.0}

\newcommand{\unaidedmayninelynxPv}{77.31\%}

\newcommand{\unaidedmayninelynxPf}{0.33\%}

\newcommand{\unaidedmayninelynxpnftotal}{742.1}
\newcommand{\unaidedmayninelynxrmse}{1799.6}

\newcommand{\unaidedmayninelynxpnfHorzTotal}{471.4}
\newcommand{\unaidedmayninelynxrmseHorz}{1713.3}

\newcommand{\unaidedmayninelynxpnfVertTotal}{209.3}
\newcommand{\unaidedmayninelynxrmseVert}{550.5}

\newcommand{\unaidedmaytwelvelynxPv}{76.94\%}

\newcommand{\unaidedmaytwelvelynxPf}{0.49\%}

\newcommand{\unaidedmaytwelvelynxpnftotal}{860.5}
\newcommand{\unaidedmaytwelvelynxrmse}{571.9}

\newcommand{\unaidedmaytwelvelynxpnfHorzTotal}{525.5}
\newcommand{\unaidedmaytwelvelynxrmseHorz}{406.1}

\newcommand{\unaidedmaytwelvelynxpnfVertTotal}{467.9}
\newcommand{\unaidedmaytwelvelynxrmseVert}{402.7}

\newcommand{\novvclynxPv}{94.86\%}

\newcommand{\singlefreqlynxPv}{90.35\%}

\newcommand{\nosbaslynxPv}{88.86\%}

\begin{document}
\title{Low-Cost Inertial Aiding for Deep-Urban Tightly-Coupled Multi-Antenna Precise GNSS}

\author{
\IEEEauthorblockN{James E. Yoder, Todd E. Humphreys} \\
\IEEEauthorblockA{\textit{Radionavigation Laboratory, The University of Texas at Austin}}
}

\maketitle

\begin{abstract}
	\ifthesis
\else 
A vehicular pose estimation technique is presented that tightly
couples multi-antenna carrier-phase differential GNSS (CDGNSS) with a low-cost
MEMS inertial sensor and vehicle dynamics constraints. This work is the first
to explore the use of consumer-grade inertial sensors for tightly-coupled urban
CDGNSS, and first to explore the tightly-coupled combination of multi-antenna
CDGNSS and inertial sensing (of any quality) for urban navigation. An unscented linearization permits ambiguity resolution using traditional integer
least squares while both implicitly enforcing known-baseline-length constraints and
exploiting the multi-baseline problem's inter-baseline correlations.
A novel false fix detection and recovery technique
is developed to mitigate the effect of conditioning the filter state on incorrect integers. When evaluated on the publicly-available TEX-CUP urban positioning dataset,
the proposed technique achieves, with consumer- and industrial-grade inertial sensors, respectively, a \baselinelynxPvShort{} and \baselinelordPvShort{}
integer fix availability, and \baselinelynxpnfHorzTotal{} cm and
\baselinelordpnfHorzTotal{} cm overall (fix and float) 95th percentile horizontal positioning error.
\fi
\end{abstract}

\newif\ifpreprint
\preprinttrue

\ifpreprint

\pagestyle{plain}
\thispagestyle{fancy}  
\fancyhf{} 
\renewcommand{\headrulewidth}{0pt}
\rfoot{\footnotesize \bf January preprint of paper submitted for review} \lfoot{\footnotesize \bf
  Copyright \copyright~2022 by James E. Yoder \\ and Todd E. Humphreys}

\else

\thispagestyle{empty}
\pagestyle{empty}

\fi


\section{Introduction}
\label{section:intro}
The rise of connected and automated vehicles has created a need
for robust globally-referenced positioning with lane-level (e.g., sub-30-cm)
accuracy \citep{reid2019localization}. Much automated ground vehicle (AGV)
research focuses on use of LIDAR and cameras for navigation, but these sensing
modalities often perform poorly in low illumination conditions or during
adverse weather such as heavy fog or snowy white-out. By contrast, positioning
techniques based on radio waves, such as automotive radar or GNSS, are robust
to poor weather and lighting conditions
\citep{narula2021radarpositioningjournal}. Recent work has found that fusing
measurements from low-cost automotive radars with inertial sensing can provide
lane-level accuracy in urban environments
\citep{narula2021radarpositioningjournal}. But radar-based positioning in a
global coordinate frame requires the time-consuming and costly production and
maintenance of radar maps.

GNSS signals provide a source of high-accuracy all-weather absolute positioning
that does not require expensive investment in systems for map production,
storage, maintenance, and dissemination. If the so-called integer ambiguities
associated with the carrier phase measurements can be correctly resolved,
carrier-phase based GNSS positioning offers exquisite accuracy.  However, GNSS
signal blockage, diffraction, and multipath effects make this family of 
techniques extremely challenging to use in urban areas. 
Carrier-phase differential GNSS (CDGNSS), whose
real-time variant for mobile platforms is commonly known as real-time kinematic
(RTK) GNSS, is a centimeter-accurate positioning technique that differences a
receiver's GNSS observables with those from a nearby fixed reference station to
eliminate most sources of measurement error
\citep[Sec. 26.3]{teunissen2017springer}. Previous work by \ifthesis
\else{this paper's authors}\fi{} probed the limits of unaided CDGNSS
in the deep urban environment, finding that
the combination of a GNSS measurement engine optimized for urban positioning 
and robust estimation techniques for outlier exclusion make CDGNSS feasible
in the deep urban environment
\citep{humphreys2019deepUrbanIts}. But the the unaided CDGNSS system described
in \cite{humphreys2019deepUrbanIts} suffers from availability gaps of up to 90
seconds in duration, making it insufficient to serve as the sole navigation sensor 
for an AGV.

A natural solution to bridging such availability gaps is to incorporate
measurements from an inertial measurement unit (IMU). These measurements are uniquely
valuable due to their invulnerability to environmental effects such as radio 
interference and weather. Combined GNSS and inertial navigation systems that
incorporate only GNSS position solutions as measurements for a downstream
navigation filter are termed \emph{loosely coupled}, whereas \emph{tightly coupled}
systems directly incorporate raw GNSS observables (pseudorange, Doppler, or
carrier phase) \citep[Sec. 28.8]{teunissen2017springer}. While both loosely- and
tightly-coupled aiding can bridge availability gaps, tightly-coupled aiding
additionally reduces these gaps' frequency and duration: the probabilistic
constraint between GNSS measurement epochs provided by the inertial sensor
increases the success rate of carrier phase integer ambiguity resolution and
makes the navigation solution observable with fewer GNSS measurements.

AGV navigation filter performance can be further improved by tightly
coupling with so-called vehicle dynamics constraints (VDCs). One such technique
exploits the natural motion constraints of four-wheeled ground vehicles,
commonly referred to as non-holonomic constraints (NHCs). A second VDC
technique infers a lack of vehicle motion by monitoring, for example, wheel
odometry ticks, or by detecting a lack of road vibration, and enforces this
constraint as a strong zero-velocity pseudo-measurement, called a zero velocity
update (ZUPT) in the literature.

This \paperOrThesis{} extends the navigation filter component of the CDGNSS system
described in \cite{humphreys2019deepUrbanIts} by tightly coupling with an
inertial sensor and with vehicle dynamics constraints, and by incorporating
measurements from multiple vehicle-mounted GNSS antennas. It also develops a
novel robust estimation technique to mitigate the effects of multipath and
allow graceful recovery from incorrect integer fixes.


\ifthesis
\else%
\subsection{Related Work}
\fi
This \ifthesis\else{subsection}\fi{} reviews relevant existing literature on urban GNSS positioning,
inertial aiding, vehicle motion constraints, and multi-antenna CDGNSS.

\ifthesis
\else%
\subsubsection{Unaided urban CDGNSS}
\fi
Performance of CDGNSS unaided by inertial sensing in urban environments has
historically been poor. Experiments in \cite{ong2009assessment} suffered from
poor availability (${<}60\%$) and large positioning errors (${>}9$m RMS) in
suburban and urban environments. A 2018 assessment of commercial CDGNSS
receivers found that no low-cost solution offered greater than $35\%$
fixed-integer solution availability in urban environments
\citep{jackson2018assessmentRtk}. \ifwiley{\cite{li2018high}}\else{Li et al. \cite{li2018high}}\fi{} achieved a 76.7\%
unaided correct integer fixing rate in urban Wuhan, China using dual-frequency
CDGNSS with a professional-grade receiver. In 2019, 
\ifwiley\else{Humphreys et al.}\fi{} \cite{humphreys2019deepUrbanIts} achieved an unaided correct integer fix
rate of $84.8\%$ in the urban core of Austin, Texas.

\ifthesis
\else%
\subsubsection{Inertial aiding}
\fi

Tightly-coupled inertial aiding has long been employed as a method to increase
CDGNSS solution availability and robustness. Early systems built around
highly-accurate but expensive tactical-grade IMUs were capable of providing
robust positioning in dense urban areas
\citep{petovello2004benefits,scherzinger2006precise,
  zhangComparisonWithTactical2006,kennedy2006architecture}. The recent
emergence of inexpensive consumer- and industrial-grade micro-electromechanical
systems (MEMS) inertial sensors has led to a new chapter of research in
low-cost inertial aiding for urban CDGNSS.


\ifwiley\else{In 2018, Li et al.}\fi{} \cite{li2018high} demonstrated that tight coupling of
single-antenna professional-grade GNSS measurements with an industrial-grade
MEMS IMU increased the integer fix availability of single-frequency CDGNSS from
44.7\% to 86.1\% on a test route in urban Wuhan, China. However, the authors
did not provide the GNSS dataset, information on the incorrect integer fix
rate, or a full error distribution, making these results difficult to assess.

This \paperOrThesis{}, in contrast, is the first to demonstrate an increased CDGNSS
integer fix rate in an urban environment via tightly coupling with a
\emph{consumer-grade} inertial sensor. Furthermore, it incorporates vehicle
dynamics constraints and multiple vehicular GNSS baselines. The system's
performance is evaluated on a publicly-available urban positioning dataset,
allowing for head-to-head comparison of techniques by the urban positioning
research community.

\ifthesis
\else%
\subsubsection{Tightly-coupled urban PPP}
\fi

One disadvantage of CDGNSS is that it requires observations from a nearby base
station to eliminate modeling errors (e.g., for atmospheric delays or satellite
clocks and orbits) common to both the base station (the reference) and the
vehicle (the rover). Short-baseline CDGNSS, which offers the greatest
robustness against urban multipath \citep{murrian2016denseNetworkPlans}, is
limited to reference-rover baseline lengths below approximately 10 km
\citep{odijk2002fast}.
To avoid the requirement for a nearby base station, attention has recently
focused on extending precise point positioning (PPP), which is based on precise
orbit, clock, and atmospheric corrections, to urban areas by tightly coupling
with inertial sensors.

Rabbou et al. in 2015 explored tight coupling of PPP with a tactical-grade
inertial sensor in mostly open-sky conditions with simulated GNSS outages,
achieving centimeter accuracy \citep{abd2015tightly}. \ifwiley{}\else{References}\fi{}
\cite{gao2017tightly} and \cite{vana2021low} extended tightly-coupled PPP to
industrial-grade MEMS inertial sensors in highway and suburban environments.
More recently, \cite{elmezayen2021ultra} demonstrated tightly-coupled PPP using
both a geodetic-grade and a low-cost GNSS receiver and an industrial-grade MEMS
sensor along an urban route in downtown Toronto, Canada, but only achieved
meter-level accuracy when using the low-cost GNSS receiver. A drawback of
PPP-based positioning is that the aforementioned results all required a roughly
10-minute convergence period before producing an accurate navigation
solution. Short-baseline CDGNSS positioning with a modern multi-frequency,
multi-constellation receiver, by contrast, typically yields instantaneous
initialization.

\ifthesis
\else%
\subsubsection{Vehicle dynamics constraints}
\fi

Recent research has also explored the tight coupling of CDGNSS measurements
with vehicle dynamics constraints. \ifwiley{}\else{Nagai et al.}\fi{} \cite{nagai2021evaluating}
found in a simulation study using a realistic 3D map of an urban environment
that a tightly-coupled CDGNSS system using GPS only could feasibly provide
high-integrity decimeter-level positioning when aided with vehicle-dynamics
constraints, a tactical-grade IMU, and odometry based on wheel-speed
sensors. \ifwiley{}\else{Yang et al. in }\fi{} \cite{yang2021improved} tightly coupled single-antenna
CDGNSS with non-holonomic constraints and a tactical-grade fiber-optic
IMU, but only evaluated their system under open-sky GNSS conditions with
simulated GNSS degradations.

\ifthesis
\else%
\subsubsection{Multi-antenna CDGNSS}
\fi

Use of multiple GNSS antennas on the vehicle for CDGNSS offers four
advantages. First, the full six-degree-of-freedom vehicle pose (position and
orientation) becomes instantaneously observable when CDGNSS measurements are
combined with the gravity vector as measured by an inertial sensor. With a
single GNSS antenna, the vehicle yaw is observable only over multiple epochs,
and only if the vehicle accelerates during the observations
\citep{hong2005observability}. Second, the shared reference antenna creates
redundancy in the measurement model that allows better ambiguity resolution
performance than any CDGNSS baseline taken individually
\citep{medina2020recursive}. Third, the additional set of GNSS measurements at
the second antenna provides reduced position estimation error. Fourth, a highly
effective method for GNSS spoofing detection, the multi-antenna defense
\citep{psiaki2014wrod}, can readily be implemented.

Multi-antenna GNSS has long been used for attitude determination applications
with snapshot estimation methods such as C-LAMBDA \citep{teunissen2006lambda}
and MC-LAMBDA \citep{giorgi2010carrier}, which provide globally-optimal
single-epoch maximum-likelihood solutions to the full nonlinear GNSS attitude
determination problem, and have been successfully extended to the pose
estimation case \citep{wu2020improving}. Other work has incorporated special
cases of \emph{a priori} attitude information into the nonlinear solution
process \citep{henkel2012reliable}. These snapshot methods, however, are
computationally demanding, and their extension to recursive estimation for
tight coupling with other sensors is not straightforward and remains
unexplored.


\ifwiley{}\else{Fan et al.}\fi{} \cite{fan2019precise} found that a hard constraint using an \emph{a
priori} known vehicle attitude to combine CDGNSS observations from multiple
vehicle antennas can increase ambiguity resolution and urban CDGNSS
performance. However, this method requires a highly-accurate independent source
of attitude information, such as from an expensive gyrocompass-capable
tactical-grade IMU following an initial static alignment period.

\ifwiley{}\else{Medina et al.}\fi{} \cite{medina2020recursive} proposed pose estimation based on
multiple vehicle antennas for inland waterway navigation.  This work
sidestepped the complexity of C-LAMBDA or MC-LAMBDA by linearizing the attitude
model in an extended Kalman filter (EKF) update and propagating the state with
a simple motion model. This formulation was found to increase ambiguity
resolution performance over either the positioning or attitude determination
problems taken independently. However, the authors made no attempt to
incorporate an inertial sensor or additional motion constraints.

\ifwiley{}\else{Hirokawa et al.}\fi{} \cite{hirokawa2009low} developed a multi-antenna GNSS system
for aircraft pose estimation that tightly coupled with a MEMS inertial sensor,
but only used CDGNSS for attitude measurements, relying on standard pseudorange
measurements for the estimator's position component.

\ifwiley{}\else{Henkel et al.}\fi{} \cite{henkel2020precise} tightly coupled triple-antenna CDGNSS
with an industrial-grade inertial sensor for a micro air vehicle navigation
application, but only evaluated the system's performance over a single, short
test flight in open-sky conditions, and did not compare against a ``ground
truth'' reference.

Previous work \citep{yoder2020visonFusion, humphreys2020owvrTracking} by this
\ifthesis\else{paper's authors}\fi{} explored a suboptimal ``federated filtering'' approach to the
tightly-coupled multi-antenna CDGNSS + inertial problem, additionally
incorporating monocular vision measurements in \cite{yoder2020visonFusion}.
But the approach did not properly model the multi-antenna CDGNSS measurement
update, instead resolving the position and attitude baselines separately.

\ifthesis
\else%
\subsection{Contributions}
\fi
This \paperOrThesis{} makes five contributions: 
\begin{enumerate}
\item An estimation technique that tightly couples multi-antenna CDGNSS with
  vehicle dynamics constraints and inertial measurements.  To the best of the
  authors' knowledge, this \paperOrThesis{} is the first in the open literature
  to explore the tightly-coupled combination of multi-antenna CDGNSS and
  inertial sensing for navigation in urban environments. Furthermore, it is the
  first to explore the use of \emph{consumer-grade} inertial sensors for
  tightly-coupled deep urban CDGNSS (Sections \ref{section:unscented} and
  \ref{section:filter}).

\item A novel application of the unscented transform for the multi-baseline
  CDGNSS integer ambiguity resolution and measurement update step, which widens
  the operating regime of the filter to allow significantly greater attitude
  uncertainty without suffering from the excessive integer least squares (ILS)
  failures seen by existing EKF approaches (Section \ref{section:unscented}).

\item A novel false fix detection and recovery technique that limits the degree
  to which an incorrectly-resolved integer ambiguity can corrupt the
  tightly-coupled CDGNSS estimator's state (Section \ref{section:false_fix}).
  
  
  
\item Demonstration of state-of-the-art deep urban CDGNSS performance,
  achieving, by tightly coupling with consumer-grade and industrial-grade
  inertial sensors, respectively, a \baselinelynxPvShort{} and
  \baselinelordPvShort{} integer fix availability, and
  \baselinelynxpnfHorzTotal{} cm and \baselinelordpnfHorzTotal{} cm overall
  (fix and float) 95th percentile horizontal positioning error on the
  publicly-available TEX-CUP urban positioning dataset \citep{narula2020texcup}
  (Sections \ref{section:setup} to \ref{section:baseline_perf}).
  
\item A detailed evaluation and breakdown of the positioning and ambiguity
  resolution performance contribution of various sensors and algorithmic
  components (Section \ref{section:alternate_configs}).
\end{enumerate}

      
\section{Coordinate and Notation Conventions}
\label{section:notation}
\subsection{Vector notation, sensor platform, and coordinate frames}
Superscripts indicate the coordinate frames associated with vectors and
rotation matrices. For example, $\vec r w$ denotes a vector $\vb r$ expressed
in the \frame w frame, and $\R w b$ denotes a rotation matrix that converts
vectors from their representation in the \frame b frame to their representation
in the \frame w frame, i.e., $\vec r w = \R w b \vec r b$.

The sensor platform described in this \paperOrThesis{} and used in the evaluation in
Section \ref{section:results} is the University of Texas \emph{Sensorium}
\citep{narula2020texcup}, a roof-mounted vehicular perception platform
incorporating multiple grades of inertial sensor, two GNSS antennas (denoted
\emph{primary} and \emph{secondary}), stereo cameras, and three automotive
radars. Only the inertial sensors and GNSS antennas are used in this work.

\begin{figure}[h]
  	\ifthesis\else
	\fontsize{10}{10}\selectfont
	\fi
	\centering
	\def\svgwidth{\ifthesis\else{1.0\columnwidth}\fi}
	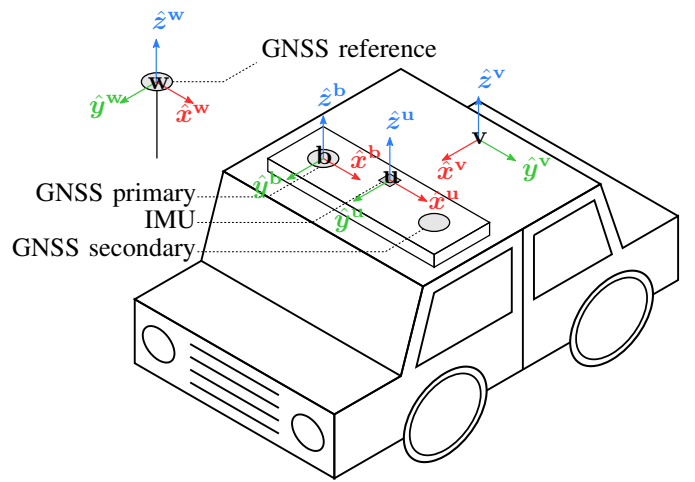
	\caption{Diagram of relevant University of Texas Sensorium coordinate frames.}
	\label{fig:frames}
\end{figure}

Several coordinate frames are referenced in this \paperOrThesis{}:
\begin{itemize}
	\item[\frame u:]{The \emph{IMU frame} is centered at and aligned with the IMU
		accelerometer triad.}
	\item[\frame b:]{The \emph{body frame} has its origin at the phase center of
		the Sensorium's primary GNSS antenna. Its $x$ axis points towards the phase
		center of the secondary antenna, its $y$ axis is aligned with the boresight
		vector of the primary antenna, and its $z$ axis completes the right-handed
		triad.}
	\item[\frame v:]{The \emph{vehicle frame} is a body-fixed frame, centered at the
		vehicle's center of rotation as determined by an offline calibration using
		GNSS and IMU data. Its $x$ axis points in the direction of vehicle travel
		with no steering angle deflection, its $z$ axis points upwards, and its $y$
		axis completes the right-handed triad.}
	\item[\frame w:]{The \emph{world frame} is a fixed geographic East-North-Up
		(ENU) frame, with its origin at the phase center of the reference GNSS
		antenna, which is located at a fixed base station with known coordinates.}
\end{itemize}
Fig. \ref{fig:frames} shows the relationships between these frames.

\subsection{State representation and error-state filtering}
The tightly-coupled navigation estimator described in this work is an unscented
Kalman filter (UKF) that recursively fuses inertial measurements,
double-difference GNSS pseudorange and carrier phase measurements, and vehicle
dynamics pseudo-measurements. The estimator's state at epoch $k$ is given by
the ordered set
\[ \vb x_k = \left(\vecs r w k, \vecs v w k, \Rs w b k, \vecs b u {\mathrm ak},
\vecs b u {\mathrm gk} \right) \] where $\vec r w \in \mathbb{R}^3$ is the
position of the \frame u frame origin in the \frame w frame;
$\vec v w \in \mathbb{R}^3$ is the velocity of the \frame u frame origin
relative to the \frame w frame, expressed in the \frame w frame;
$\R w b \in \mathrm{SO(3)}$ is the attitude of the \frame b frame relative to
the \frame w frame; and
$\vecs b u {\mr{a}}, \vecs b u {\mr{g}} \in \mathbb{R}^3$ are the IMU's
accelerometer and gyro biases, respectively.

Because the set of 3D rotations, which can be represented using the special
orthogonal group $\mathrm{SO(3)}$, is not a vector space, certain adaptations
are needed to the typical Kalman filter equations to properly account for its
manifold structure. A popular and well-founded method is to use a full,
nonsingular attitude parameterization (e.g., quaternions or rotation matrices)
in the filter state, but to express uncertainties, velocities, and small
increments of the state using a minimal vector space parameterization that is
local to the nominal state. Examples of this approach appear in the robotics
literature as as the error state Kalman filter \citep{sola2017quaternion}, and
in the aerospace literature as the multiplicative EKF
\citep{markley2004multiplicative}.

This \paperOrThesis{} adopts the conventions and notation of \ifwiley{}\else{Sol\'a et
al.}\fi{} \cite{sola2018micro}, which appeals to Lie theory to unify and generalize
the various methods so that the filtering equations are agnostic to the
specific choice of attitude parameterization. The filter state $\vb x_k$ is a
point on the composite manifold
$\mathcal{X} \equiv \mathbb{R}^3 \times \mathbb{R}^3 \times \mathrm{SO(3)}
\times \mathbb{R}^3 \times \mathbb{R}^3$, which has $N_{x} = 15$ independent degrees of freedom. A state increment $\delta \vb x_k$ is
defined on the tangent space of $\mathcal X$ at $\vb x_k$, which can be parameterized with the vector
space $\mathbb{R}^{N_x}$. These spaces are related using the operators
$\oplus : \mathcal X \times \mathbb{R}^{N_x} \rightarrow \mathcal X$ and
$\ominus : \mathcal X \times \mathcal X \rightarrow \mathbb{R}^{N_x}$, which
correspond to standard addition and subtraction for vector-valued components of
$\vb x_k$ and to more complex operations for the attitude component:
\begin{align*}
\vb x_k \oplus \delta \vb x_k \define &{}
\begin{bmatrix}
\vecs r w k + \delta \vecs r w k \\
\vecs v w k + \delta \vecs v w k \\
\Rs w b k \circ \Exp{\left(\delta \Rs w b k\right)} \\
\vecs b u {\mathrm ak} + \delta \vecs b u {\mathrm ak} \\
\vecs b u {\mathrm gk} + \delta \vecs b u {\mathrm gk}
\end{bmatrix}
\in \mathcal X \\
\vb x_j \ominus \vb x_k \define &{}
\begin{bmatrix}
\vecs r w j - \vecs r w k \\
\vecs v w j - \vecs v w k \\
\Log{\left(\Rs w b j \circ {\Rs w b k}^{-1}\right)} \\
\vecs b u {\mathrm aj} - \vecs b u {\mathrm ak} \\
\vecs b u {\mathrm gj} - \vecs b u {\mathrm gk}
\end{bmatrix}
\in \mathbb{R}^{N_x}
\end{align*}
where $\circ$ denotes rotation composition,
$\Exp : \mathbb{R}^{3} \rightarrow \mathrm{SO(3)}$, and
$\Log : \mathrm{SO(3)} \rightarrow \mathbb{R}^{3}$. The estimator's attitude
parameterizations are rotation matrices and 3-1-2 Euler angles for the state
and tangent space, respectively, leading to the definitions
\begin{flalign*}
&\Exp :
\begin{bmatrix} \phi \\ \theta \\ \psi \end{bmatrix}
\mapsto
\begin{bmatrix} 
\c\psi\c\theta - \s\phi\s\psi\s\theta & \c\theta\s\psi+\c\psi\s\phi\s\theta & -\c\phi\s\theta \\
-\c\phi\s\psi & \c\phi\c\psi & \s\phi \\
\c\psi\s\theta+\c\theta\s\phi\s\psi & \s\psi\s\theta-\c\psi\c\theta\s\phi & \c\phi\c\theta
\end{bmatrix}& \\
&\Log :
\begin{bmatrix}
R_{11} & R_{12} & R_{13} \\
R_{21} & R_{22} & R_{23} \\
R_{31} & R_{32} & R_{33} 
\end{bmatrix} 
\mapsto
\begin{bmatrix}
\arcsin R_{23} \\
\arctan \frac {R_{33}}{-R_{13}} \\
\arctan \frac {R_{22}}{-R_{21}}
\end{bmatrix}&
\end{flalign*}
where $\arctan$ denotes the 4-quadrant arctangent (i.e., \texttt{atan2}),
$\c x$ denotes $\cos(x)$, and $\s x$ denotes $\sin(x)$.  These attitude
parameterizations could, of course, easily be substituted with alternate
parameterizations such as unit quaternions and axis-angle rotation vectors,
with appropriate redefinition of the $\Exp$ and $\Log$ maps following
\cite{sola2018micro}.

Probabilistic beliefs under this framework are taken as Gaussian distributions
over the tangent space.  Let $Z^k$ be the set of all measurements up to time
$k$.  Then with $\vb x_k$ denoting the true system state at $k$, define the
\emph{a priori} state $\vbb x_k$, its error covariance $\vbb P_k$, the \emph{a
  posteriori} state $\vbh x_k$, and its error covariance $\vbh P_k$, as
follows:
\begin{align*}
\vbb x_k &\define   \E{\vb x_k| Z^{k-1}} \in \mathcal X  \\
\vbb P_k &\define \E{(\vb x_k \ominus \vbb x_k)(\vb x_k \ominus \vbb x_k)^\tr
	| Z^{k-1}} \in \mathbb{R}^{N_x \times N_x}\\
\vbh x_k &\define  \E{\vb x_k| Z^k} \in \mathcal X \\
\vbh P_k &\define \E{(\vb x_k \ominus \vbh x_k)(\vb x_k \ominus \vbh x_k)^\tr
	| Z^k}\in \mathbb{R}^{N_x \times N_x}
\end{align*}
The remarkable feature of this notational framework is that the typical Kalman
filtering equations can be adapted for on-manifold estimation by simply
replacing $+$ and $-$ with the $\oplus$ and $\ominus$ operators as needed.

\section{An Unscented Multi-Baseline CDGNSS Measurement Update}
\label{section:unscented}

\subsection{CDGNSS measurement model}
\begin{figure}[t]
  \ifthesis\else
  	\fontsize{10}{10}\selectfont
  \fi
  \centering
	\def\svgwidth{\ifthesis\else{1.0\columnwidth}\fi}
  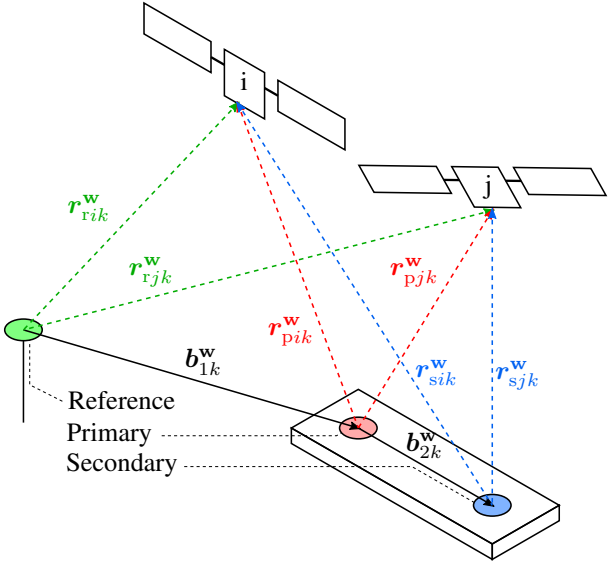
  \caption{Baseline and antenna-to-satellite vectors of the multi-antenna
    CDGNSS measurement model, for pivot satellite $i$ and non-pivot satellite
    $j$. $\vecs r w {\mr rnk}$, $\vecs r w {\mr pnk}$, and $\vecs r w {\mr snk}$ refer to vectors pointing from the reference, primary, and secondary GNSS antennas, respectively, to the antenna phase center of GNSS satellite $n$.}
  \label{fig:diagram}
\end{figure}

At each GNSS measurement epoch, the estimator ingests $N_k$ pairs of
double-difference (DD) GNSS observables, each pair composed of a pseudorange
and a carrier phase measurement, across all baselines. The baselines and
relevant relative position vectors are shown in Fig. \ref{fig:diagram}.  The
measurement vector at epoch $k$ is
\begin{align*}
  \vb z_{\mr{g}k}
  \define \left[
  \vb \rho_{1k}^\tr, \vb \phi_{1k}^\tr, \vb \rho_{2k}^\tr, \vb \phi_{2k}^\tr
  \right]^\tr \in \mathbb{R}^{2N_k}
\end{align*}	
where $\vb \rho_{mk}$ and $\vb \phi_{mk}$ are vectors of double-difference
pseudorange and carrier phase measurements, both in meters, for baseline
$m \in \{1,2\}$ at epoch $k$. The measurement $\vb z_{\mr gk}$ is a function of
the state $\vb x_k$, the integer ambiguity vector
$\vb n_k \in \mathbb{Z}^{N_k}$, and zero-mean white Gaussian measurement noise
$\vb \epsilon_{\mr gk}$ \citep{psiaki2007modeling}:
\begin{equation}
  \begin{aligned}
    \vb z_{\mr gk} =
    \vb h_{\mr gk}\left(\vb b\left(\vb x_k\right), \vb n_k\right) + \vb
    \epsilon_{\mr gk}, \quad
    \vb \epsilon_{\mr gk} \sim  \mathcal{N}\left(\vb 0, \vb \Sigma_{\mr gk}\right)
  \end{aligned}
\end{equation}
The function $\vb b(\vb x_k)$ relates the baseline vectors $\vecs b w {1k}$ and
$\vecs b w {2k}$ shown in Fig. \ref{fig:diagram} to the position and attitude
components of the state:
\begin{equation}
  \label{eq:baselines}
  \begin{aligned}
    \vb b(\vb x_k) \define 
    \begin{bmatrix}
      \vecs b w {1k} \\
      \vecs b w {2k} 
    \end{bmatrix} = 
    \begin{bmatrix}
      \vecs r w k + \Rs w b k \left(\vecs r b {\mr p} - \vecs r b {\frame u} \right)			\\
      \Rs w b k \left(\vecs r b {\mr s} - \vecs r b {\mr p} \right)
    \end{bmatrix}
  \end{aligned}
\end{equation}
Here, $\vecs r b {\frame u}$, $\vecs r b {\mr p}$, and
$\vecs r b {\mr s}$ are the body-frame positions of the IMU, primary GNSS
antenna, and secondary GNSS antenna, respectively. Under this formulation, the
known length of $ \vecs b w {2k} $ serves as an implicit constraint on integer
ambiguity resolution due to the parameterization of $\vecs b w {2 k}$ as a
function solely of attitude.

Importantly, the off-diagonal blocks of the measurement noise covariance matrix
$\vb \Sigma_{\mr gk}$ are nonzero because baselines 1 and 2 share a GNSS antenna. By
consuming the GNSS measurements for all baselines in a single update, this
correlation is exploited and typically yields a higher integer fix success rate
than either baseline taken individually
\citep{medina2020recursive}. Additionally, the vehicle attitude $\Rs w b k$ is
often known \emph{a priori} to sub-degree precision, providing a tight
constraint on all 3 degrees of freedom of $\vecs b w {2 k}$, which both
strengthens the combined integer model \citep{fan2019precise} and increases
positioning accuracy, since the off-diagonal blocks of $\vb \Sigma_{\mr g k}$ encode
the sensitivity of GNSS measurements on the secondary vehicle antenna to
$\vecs r w k$.
	
\subsection{Linearization}
The function $\vb h_{\mr gk}(\vb b, \vb n_k)$ is accurately modeled as linear
due to the extreme distance to GNSS satellites relative to the CDGNSS baseline
lengths:
\begin{align*}
  \vb h_{\mr gk}(\vb b(\vb x_k), \vb n_k) =
  \begin{bmatrix}
    \vb \rho_{1 k} \\ \vb \phi_{1 k} \\
    \vb \rho_{2 k} \\ \vb \phi_{2 k} 
  \end{bmatrix} = {}&{}
                      \begin{bmatrix*}[l]
                        \vb G_{1k}\vecs b w {1} \\
                        \vb G_{1k}\vecs b w {1} + \vb \Lambda_1 \vb n_{1k} \\
                        \vb G_{2k}\vecs b w {2} \\
                        \vb G_{2k}\vecs b w {2} + \vb \Lambda_2 \vb n_{2k}\\
                      \end{bmatrix*}
\end{align*}
Here, $\vb n_{m k} \in \mathbb{Z}^{N_{m k}}$ is the vector of carrier-phase
integer ambiguities for baseline $m$,
$\vb n_k = \begin{bmatrix}\vb n_{1k}^\tr, \vb n_{2k}^\tr\end{bmatrix}^\tr$, and
$\vb\Lambda_m$ is a diagonal matrix composed of the wavelengths in meters of
each DD carrier phase measurement. The geometry matrix $\vb G_{mk}$ is defined
as
\begin{align*}
  \vb G_{mk} \define
  \begin{bmatrix}
    \left(\vbh r^{\frame w}_{ik} - \vbh r^{\frame w}_{1k}\right)^\tr \\
    \vdots \\
    \left(\vbh r^{\frame w}_{ik} - \vbh r^{\frame w}_{N_{mk}k}\right)^\tr
  \end{bmatrix}
\end{align*}
for pivot satellite $i$ and non-pivot satellites $1$ to $N_{mk}$, where
$\vbh r^{\frame w}_{jk}$ denotes a unit vector in the \frame w frame directed
from a GNSS antenna to GNSS satellite $j$ of baseline $m$ at epoch $k$. Under
the small-angle approximation, these unit vectors are assumed to be
approximately equal for all receiver antennas involved:
\begin{align*}
  \vbh r^{\frame w}_{ik} 
  \approx \frac{\vb r^{\frame w}_{\mr r i k}}{\snorm{ \vb r^{\frame w}_{\mr r i k}}}
  \approx \frac{\vb r^{\frame w}_{\mr p i k}}{\snorm{ \vb r^{\frame w}_{\mr p i k}}}
  \approx \frac{\vb r^{\frame w}_{\mr s i k}}{\snorm{ \vb r^{\frame w}_{\mr s i k}}}
\end{align*}

The nonlinearity of (\ref{eq:baselines}) due to the manifold structure of
vehicle attitude $\Rs w b k$ is nontrivial. Optimal snapshot estimators for the
nonlinear GNSS attitude problem, such as C-LAMBDA \citep{teunissen2006lambda}
and MC-LAMBDA \citep{giorgi2010carrier}, have been studied, but these estimators
are computationally expensive and their extension to recursive filtering does
not seem straightforward. Instead, extant Kalman filtering-based multi-baseline
CDGNSS estimators \citep{medina2020recursive,henkel2020precise} typically
linearize the baseline measurement model about the current state estimate with
a simple first-order Taylor expansion in order to perform integer ambiguity
resolution with a standard ILS solver such as the well-known LAMBDA method
\citep{teunissen1995LAMBDA_journal}. This is essentially an extension of the
LC-LAMBDA method described by \ifwiley{}\else{Teunissen et al. in}\fi{} \cite{teunissen2009extent} to
the multi-baseline recursive estimation case. As demonstrated in
\cite{teunissen2009extent}, this method performs poorly for ultra-short (length
$\lesssim1$m) baselines when the \emph{a priori} attitude estimate and the
pseudorange measurements cannot together offer a sufficiently accurate estimate
about which to linearize.
	
\subsubsection{Alternatives}
Recursive Bayesian estimation of $\vb x_k$ requires finding the distribution of
$\vb b(\vb x_k)$ given a Gaussian prior for $\vb x_k$ with mean $\vbb x_k$ and
covariance $\vbb P_k$. While the true distribution of $\vb b(\vb x_k)$ is
nontrivial, it is desirable to approximate it as Gaussian to enable ambiguity
resolution with standard ILS techniques, which are computationally efficient
and well understood. Dropping the $k$ subscripts for notational clarity, such
an approximation produces a joint Gaussian distribution over $\vb x$ and
$\vb b(\vb x)$:
\begin{align*}
  \begin{bmatrix}\vb x\\\vb b(\vb x)\end{bmatrix} \sim 
  \mathcal{N}\left(
  \begin{bmatrix}\vbb x\\\vbb b\end{bmatrix},
  \begin{bmatrix}
    \vbb P & \vb P_{xb} \\
    \vb P_{xb}^\tr & \vb P_{bb}
  \end{bmatrix}
                     \right)
\end{align*}
This can be parameterized in terms of mean $\vbb b$, Jacobian $\vb H_{\mr b}$,
and additional baseline uncertainty $\vbs \Sigma_{\mr b}$ that accounts for errors
due to linearization:
\begin{equation}
\label{eq:pbb_pbx}
\begin{aligned}
  \vb P_{bb} \define {}&{} \vb H_{\mr b} \vbb P \vb H_{\mr b}^\tr + \vbs \Sigma_{\mr b}\\
  \vb P_{xb} \define {}&{}  \vbb P \vb H_{\mr b}^\tr
\end{aligned}
\end{equation}
The first-order Taylor expansion scheme used by extant multi-baseline CDGNSS
Kalman filters to obtain these parameters can be described as in Algorithm
\ref{algo:ekf}.
\begin{inlinealgo}{linearizeEkf}
  \label{algo:ekf}
  \Input{$\vbb x$, $\vbb P$}
  \Output{$\vbb b$, $\vb H_{\mr{b}}$, $\vbs \Sigma_{\mr{b}}$}
  
  $\vbb b = \begin{bmatrix}
    \vecs b w {1} \left(\vbb x\right) \\			
    \vecs b w {2} \left(\vbb x\right) \\
  \end{bmatrix}$
  
  $\vb H_{\mr{b}} = 
  \left.
    \renewcommand*{\arraystretch}{1.5}
    \begin{bmatrix}
      \pdv{\vecs b w 1 (\vb x)}{\vb x} \\
      \pdv{\vecs b w 2 (\vb x)}{\vb x}
    \end{bmatrix}
  \right\rvert_{\vbb x}
  $
  
  $\vbs \Sigma_{\mr b} = \vb 0$
\end{inlinealgo}

An alternative approach is to approximate the distribution of $\vb b(\vb x)$
using a deterministic sampling technique such as the unscented transform (UT),
which is used in the UKF \citep{julier2004UKF}. The UT infers the probability
distribution of a transformed Gaussian by transforming a set of weighted
``sigma points'' through the nonlinearity and evaluating the statistics of the
transformed points. An implementation of the UT is shown in Algorithm
\ref{algo:ukf}.
\begin{inlinealgo}{linearizeUkf}
    \label{algo:ukf}
    \Input{$\vbb x$, $\vbb P$}
    \Output{$\vbb b$, $\vb H_{\mr{b}}$, $\vbs \Sigma_{\mr{b}}$}
    $\alpha = 0.001,\; \kappa = 0$
    
    $\lambda = \alpha^2 \left(N_x + \kappa\right) - N_x$
    
    $\vb S = \begin{bmatrix}
      \vb s_1, \vb s_2, \dots, \vb s_{N_x}
    \end{bmatrix} = \chol{\vbb P}^\tr$
    
    $\vbb x^{(0)} = \vbb x$
    
    $\vbb b^{(0)} = \begin{bmatrix} 
      \vecs b w 1 \left(\vbb x^{(0)}\right) \\
      \vecs b w 2 \left(\vbb x^{(0)}\right) \\
    \end{bmatrix} $ 
    
    $w_m^{(0)} = \frac{\lambda}{N_x + \lambda}$
    
    $w_c^{(0)} = \frac{\lambda}{N_x + \lambda} + 1 - \alpha^2 + \beta$
   
    	\For{$i \in [1, 2N_x]$}{\hbox {\ifthesis\else \vrule \fi \vtop{
$
          \vbb x^{(i)} = 
          \begin{cases}
            \vbb x^{(0)} \oplus  \sqrt{N_x+\lambda} \,\vb s_i & i \in [1, N_x] \\
            \vbb x^{(0)} \oplus  -\sqrt{N_x-\lambda} \,\vb s_i & i \in [N_x+1,2N_x]
          \end{cases}$
          
          $\vbb b^{(i)} = \begin{bmatrix} 
            \vecs b w 1 \left(\vbb x^{(i)}\right) \\
            \vecs b w 2 \left(\vbb x^{(i)}\right) \\
          \end{bmatrix}$
          
          $w_m^{(i)} = w_c^{(i)} = \frac{1}{2(N_x+\lambda)}$
        }}
    }
    
    $\vbb b = \displaystyle\sum_{i=0}^{2N_x}{ w_m^{(i)} \vbb b^{(i)}} $
    
    \vspace{5pt}
    $ \begin{bmatrix}
      \vbb P_{xx} 	& \vbb P_{xb} \\ 
      \vbb P_{xb}^\tr & \vbb P_{bb} 
    \end{bmatrix} = 
    \displaystyle\sum_{i=0}^{2N_x}{\left( w_c^{(i)} 
        \begin{bmatrix}
          \vbb x^{(i)} \ominus \vbb x^{(0)} \\
          \vbb b^{(i)} - \vbb b 
        \end{bmatrix}
        \begin{bmatrix}
          \vbb x^{(i)} \ominus \vbb x^{(0)} \\
          \vbb b^{(i)} - \vbb b 
        \end{bmatrix}^\tr
      \right)}$
    
    \vspace{5pt}
    %
    
    $ \vb H_{\mr b} =  \left(\vbb P_{xx}^{-1} \vbb P_{xb}\right)^\tr$
    
    $ \vbs \Sigma_{\mr b} = \vbb P_{bb} - \left(\vb H_{\mr b} \vbb P_{xx}\vb H_{\mr b}^\tr \right)$%
\end{inlinealgo}
\begin{figure}[h]
  \centering
  \includegraphics[width={\ifthesis\else\columnwidth\fi}]{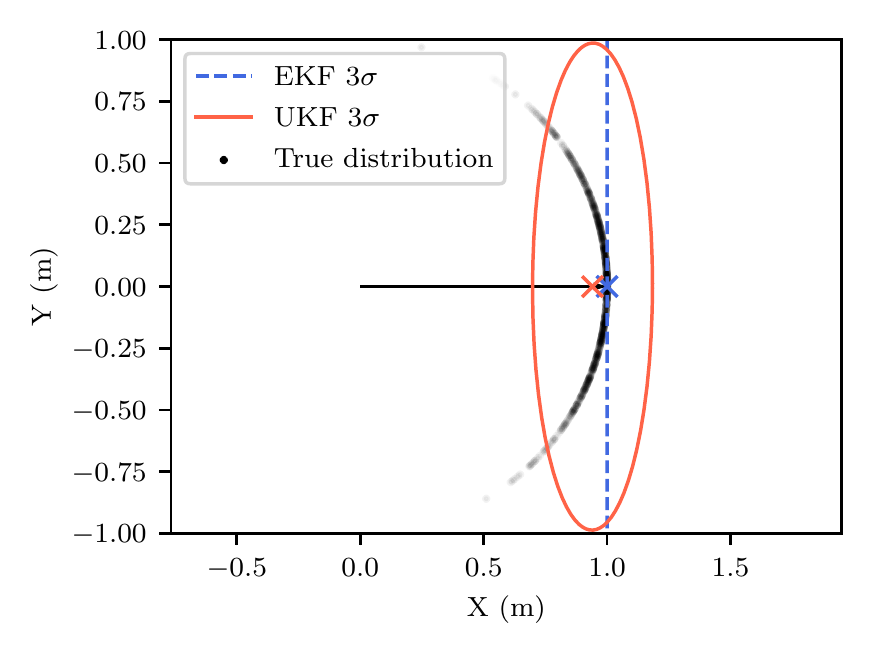}
  \caption{Simplified 2-dimensional example of Gaussian distributions produced
    by \texttt{linearizeEkf} and \texttt{linearizeUkf} for a single constrained
    baseline of length 1 m with a Gaussian prior over the attitude angle. The
    lines represent the cost contours in the position domain associated with
    each Gaussian distribution at the $3\sigma$ likelihood; the $\times$
    symbols represent the means. The EKF distribution is infinitely thin, as
    EKF linearizations only consider baseline vectors on a plane tangent to the
    true spherical distribution at the \emph{a priori} estimate, causing ILS
    failures when the \emph{a priori} uncertainty is large enough that the
    sphere significantly diverges from the tangent plane.}
  \label{fig:linearization_diagram}
\end{figure}

A simplified 2-dimensional example of the two linearization schemes for a
single constrained baseline is shown in Fig. \ref{fig:linearization_diagram}.
The UT yields an approximate Gaussian distribution over the baseline vector
that more closely matches the true mean and covariance of $\vb b(\vb x)$.  For
this reason, ILS-based ambiguity resolution has a higher success rate when
linearization is based on \texttt{linearizeUkf} rather than
\texttt{linearizeEkf} under large \emph{a priori} attitude uncertainty, as will
be demonstrated in Section \ref{section:evaluation_of_unscented}.
		
\subsubsection{UKF measurement update}
A vector $\vb z_{\mr gk}$ of DD GNSS observables is ingested by the estimator
at epoch $k$. Linearizing about the \emph{a priori} state estimate $\vbb x_k$
\begin{equation*}
  \left[\vbb b_k, \vb H_{\mr bk}, \vbs \Sigma_{\mr bk}\right] = 
  \texttt{linearizeUkf}\left(\vbb x_k, \vbb P_k\right) \\
\end{equation*}
yields the following approximation of the measurement model for innovations
vector
$\vb \nu_{\mr{g} k} \define \vb z_{\mr{g}k} - \vb h_{\mr{g}k}\left(\vbb b_k,
  \vb 0\right)$:
\begin{align*}
  \vb \nu_{\mr{g k}} = &
                           \underbrace{
                           \begin{bmatrix}
                             \vb G_{1k} 					& \vb 0 	\\
                             \vb G_{1k} 	& \vb 0 	\\
                             \vb 0                 				& \vb G_{2k} \\
                             \vb 0                 				& \vb G_{2k} \\
                           \end{bmatrix}
  \vb H_{\mr bk}
  }_{\vb H_{\mr{r}k}}
  \delta \vb x_k + 
                        \underbrace{
			\begin{bmatrix}
                          \vb 0   		& \vb 0 \\
                          \vb \Lambda_1   & \vb 0 \\
                          \vb 0           & \vb 0 \\
                          \vb 0           & \vb \Lambda_2
			\end{bmatrix}
                                            }_{\vb H_{\mr nk}}
                                            \underbrace{ 
                                            \begin{bmatrix}
                                              \vb n_{1k} \\ \vb n_{2k}
                                            \end{bmatrix}
  }_{\vb n_k} \\
  & + \vb \epsilon_{\mr{g}k} + \vbs \epsilon_{\mr{g}k} \\
  = & ~\vb H_{\mr rk} \delta \vb x_k + 
        \vb H_{\mr nk} \vb n_k +
        \vb \epsilon_{\mr gk} + \vbs \epsilon_{\mr{g}k} \\
                         &	\vbs \epsilon_{\mr{g}k} \sim  \mathcal{N}\left(\vb 0, \vb H_{\mr bk} \vbs \Sigma_{\mr bk} \vb H_{\mr bk}^\tr \right)
\end{align*}
Here the state estimate error vector
$\delta \vb x_k \define \vbb x_k \ominus \vb x_k$ is expressed in the tangent
space of $\mathcal{X}$ at $\vbb x_k$, and $\vbs \epsilon_{\mr gk}$ represents
additional measurement error caused by approximation of
$\vb b \left(\vb x_k\right)$.

\subsection{Square-root formulation}
\label{section:sr}
The CDGNSS measurement update can be cast in square-root form for greater
numerical robustness and algorithmic clarity \citep{psiaki2005relative}.  Given
$\vb \nu_{\rm gk}$, $\vb H_{\rm rk}$, $\vb H_{\rm nk}$, $\vbb x_k$, and
$\vbb P_k$, the measurement update can be defined as finding $\delta \vb x_k$
and $\vb n_k$ to minimize the cost function
\begin{align*}
  J_k(\delta \vb x_k, \vb n_k) =
  \norm{
    \vb \nu_{\mr gk} 
    - \vb H_{\mr rk}\delta \vb x_k 
    - \vb H_{\mr nk} \vb n_k
  }^2_{\vb \Sigma_{k}^{-1}} +
  \norm{
    \vb \delta x_k
  }^2_{\vbb P_k^{-1}}
\end{align*}
where
$\vb \Sigma_{k} = \vb \Sigma_{\mr gk} + \vb H_{\mr b} \vbs \Sigma_{\mr bk} \vb
H_{\mr b}^\tr$. The vector cost components can be normalized by left
multiplying with square-root information matrices based on Cholesky
factorization $\vb R_{\mr gk} = \chol{\vb \Sigma_{k}{}^{-1}}$,
$\vbb R_{xxk} = \chol{\vbb P_k^{-1}}$:
\begin{align*}
  J_k(\delta &\vb x_k, \vb n_k) \\ = &\norm{
    \begin{bmatrix}
      \vb 0 \\ \vb R_{\mr gk} \vb \nu_{\mr gk}
    \end{bmatrix} -
    \begin{bmatrix}
      \vbb R_{xxk} \\ \vb R_{\mr gk} \vb H_{\mr rk} 
    \end{bmatrix}
    \delta \vb x_k -
    \begin{bmatrix}
      \vb 0 \\ \vb R_{\mr gk} \vb H_{\mr rk}
    \end{bmatrix}
    \vb n_k
  }^2 \\ 
  = &\norm{
    \vb \nu_k' - 
    \begin{bmatrix}
      \vb H_{\mr rk}' \\
      \vb H_{\mr nk}'
    \end{bmatrix}
    \begin{bmatrix}
      \delta \vb x_k \\ \vb n_k
    \end{bmatrix}
  }^2
\end{align*}
The cost $J_k$ can be decomposed via QR factorization
\begin{align*}
  \left[\vb Q_k, \vb R_k \right] = \mathtt{qr}\left(
  \begin{bmatrix}
    \vb H_{\mr rk}' \\
    \vb H_{\mr nk}'
  \end{bmatrix}
  \right)
\end{align*}
where matrix $\vb Q_k$ is orthogonal and $\vb R_k$ is upper triangular.
Because $\vb Q_k$ is orthogonal, the components of $J_k$ inside the norm can be
left-multiplied by $\vb Q_k^\tr$ without changing the cost, and $J_k$ can be
decomposed into 3 terms:
\begin{align}
  \label{eq:sr_cost} 
   J_k(\delta &\vb x_k, \vb n_k)  =  \norm{
               \vb Q_k^\tr \vb \nu'_k -
               \vb R_k 
               \begin{bmatrix}
                 \delta \vb x_k  \\ \vb n_k
               \end{bmatrix}	
  }^2 \nonumber \\ 
  = & \norm {
      \begin{bmatrix}
        \vb \nu''_{1k} \\ \vb \nu''_{2k} \\ \vb \nu''_{3k}
      \end{bmatrix} - 
  \begin{bmatrix}
    \vb R_{xxk} & \vb R_{xnk} \\
    \vb 0      & \vb R_{nnk} \\
    \vb 0      & \vb 0
  \end{bmatrix}
                 \begin{bmatrix}
                   \delta \vb x_k \\ \vb n_k
                 \end{bmatrix}
  }^2  \\  
  = & \underbrace{\norm{
      \vb \nu_{1k}'' 
      - \vb R_{xxk} \delta \vb x_k 
      - \vb R_{xnk} \vb n_k
      }^2}_{J_{1k}\left(\delta\vb x_k, \vb n_k\right)} 
      + \underbrace{\norm{ 
      \vb \nu_{2k}'' - \vb R_{nnk} \vb n_k				
      }^2}_{J_{2k}(\vb n_k)} \nonumber \\ 
    & + \underbrace{\norm {\vb \nu_{3k}''}^2}_{J_{3k}} \nonumber
\end{align}
If both the measurement model and
$\vbb R_{xxk}$ are not ill-conditioned, then $\vb R_{xxk}$ and
$\vb R_{nnk}$ are invertible. $J_{3k}$ is the irreducible cost, and, under a
single-epoch ambiguity resolution scheme, can be shown to be equal to the
normalized innovations squared (NIS) associated with the double-difference
pseudorange measurements. $J_{2k}$ is the extra cost incurred by enforcing the
integer constraint on $\vb n_k$, and can similarly be shown to be equal to the
NIS of the double-difference carrier phase measurements (again assuming
single-epoch ambiguity resolution). If $\vb n_k$ is allowed to take any real
value (the ``float solution''), $J_{2k}$ can be zeroed due to the invertibility
of $\vb R_{nnk}$. Similarly, $J_{1k}$ can be zeroed for any value of $\vb n_k$
due to the invertibility of $\vb R_{xxk}$.  The ``float solution,''
$\{\delta \vbt x_k, \vbt n_k\}$, can therefore be formed by choosing
$\delta \vbt x_k$ and $\vbt n_k$ to zero $J_{1k}$ and $J_{2k}$. Because
$\vb R_k$ is upper triangular, these values can be found by efficient
backsubstitution.  The ``fixed solution,'' $\{\delta \vbc x_k,\vbc n_k\}$, is
via an ILS solver, yielding
\begin{equation}
  \label{eq:ils_solve}
  \begin{aligned}
    \vbc n_k = {}& \arg \underset{\vb n_k \in \mathbb{Z}^{n}}{\min} \, J_{2k}(\vb n_k) \\
    \delta \vbc x_k = {}& \vb R_{xxk}^{-1} \left( \vb \nu_{1k}'' - \vb R_{xnk} \right)
  \end{aligned}
\end{equation}	
$\vb R_{xxk}$ is the \emph{a posteriori} state vector square-root information
matrix conditioned on $\vb n_k = \vbc n_k$. Therefore, if the fixed solution is
accepted (having passed validation via an integer aperture test), the \emph{a
  posteriori} state and covariance are
\begin{equation}
  \label{eq:accept_fix}
  \begin{aligned}
    \vbh x_k = {}& \vbb x_k \oplus \delta \vbc x_k \\
    \vbh P_k = {}& \left( \vb R_{xxk}^\tr \vb R_{xxk} \right)^{-1}
  \end{aligned}
\end{equation}
If instead the float solution is accepted, the \emph{a posteriori} state and
covariance are found by marginalizing over the distribution of $\vb n_k$:
\begin{equation}
  \label{eq:accept_float}
  \begin{aligned}
    \vbh x_k = {}& \vbb x_k \oplus \delta \vbt x_k \\
    \vbh P_k = {}& (\vb R_k^\tr \vb R_k)^{-1}_{[1:N_x,1:N_x]}
  \end{aligned}
\end{equation}
Here, $[1:n, 1:n]$ denotes taking the first $n$ rows and columns
of the matrix.
		
\subsection{Evaluation of Unscented Multi-Antenna Update}
\label{section:evaluation_of_unscented}

\begin{figure}[h]
	\centering
	\includegraphics[width={\ifthesis\else\columnwidth\fi}]{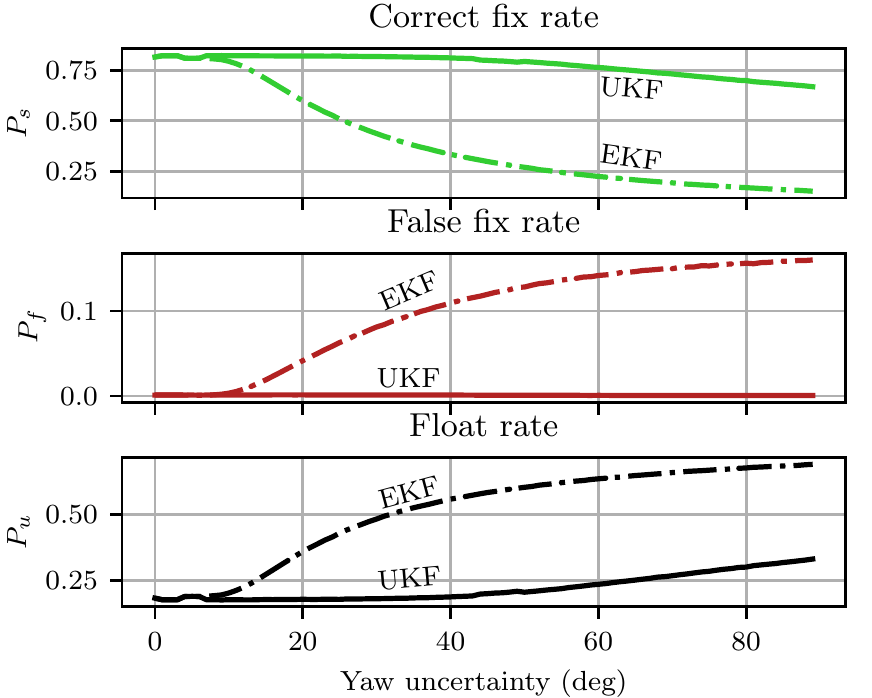}
	\caption{Integer aperture (IA) success (validated correct fix), failure
          (validated false fix), and float (failed IA validation) rates found via 
          \ifthesis
\else{Monte Carlo simulation with $10^6$
          	samples for an example multi-baseline CDGNSS measurement update with
          	varying \emph{a priori} attitude uncertainty}
          \fi. 
          \ifthesis\else{
          	UKF and EKF denote the ILS success rates for
          	models derived from \texttt{linearizeUkf} and \texttt{linearizeEkf},
          	respectively.
          	The vehicle pitch and roll
          	angles were assumed to be known to $2^\circ$ ($1\sigma$), and the
          	\emph{a priori} yaw angle uncertainty was varied from $0^\circ$ to
          	$90^\circ$ ($1\sigma$). The simulation
          	assumed GPS L1 C/A signals visible from the equator with a
          	representative satellite constellation and a $15^\circ$ elevation
          	mask angle. Two vehicle GNSS antennas were simulated, with a baseline
          	length of $1.0668$ m (equivalent to that of the Sensorium). Other
          	GNSS measurement model parameters were selected as in Table
          	\ref{table:ppengine_config}.
          }\fi
      	  The threshold function
          approximation to the fixed-failure rate distance test
          \citep{wang2015iafit} was used with fixed failure rate
          $\bar P_f = 0.01$. Linearization error causes the true failure rate
          $P_f$ of the EKF linearization to greatly exceed $\bar P_f$ as the
          yaw uncertainty increases beyond approximately $8^\circ$. In
          contrast, the integrity of the integer aperture test is maintained
          for the UKF case due to the approximate linearization error term
          $\vbs \Sigma_\mr{b}$ provided by the UKF linearization.  Note that
          $P_u = 1 - P_f - P_s$ \citep{green2018iagiab}.}
	\label{fig:linearization_montecarlo_ia}
\end{figure}

The UKF linearization of $\vb b(\vb x_k)$ yields a Gaussian prior in the
position domain that better captures the true mean and covariance of the
constrained attitude baseline than the EKF linearization. Consequently, the UKF
linearization achieves a higher ILS success rate than the EKF linearization
when attitude uncertainty is large. Fig. \ref{fig:linearization_montecarlo_ia}
demonstrates this effect by evaluating integer aperture success, failure, 
and float rates for a multi-baseline
CDGNSS measurement update via Monte Carlo simulation. For a dual-antenna
platform similar to that of the University of Texas Sensorium with loose yaw
knowledge, this effect becomes apparent as $1\sigma$ yaw uncertainty exceeds
approximately $8^\circ$, whereupon the ILS success rate with the EKF
linearization begins a rapid decline with increasing uncertainty, eventually
falling below even that of the unconstrained multi-antenna CDGNSS snapshot
estimator.

The UKF linearization expands the operating regime of the CDGNSS navigation
estimator to greater levels of attitude uncertainty than with the EKF
linearization. This effect is relevant for low-cost urban CDGNSS: while even a
low-cost accelerometer can provide pitch and roll angles with degree-level
accuracy, it is desirable to tolerate large yaw uncertainty, as may occur, for
example, following a long GNSS outage in a parking garage. Upon emerging from
such an outage, a system using an EKF linearization may require
re-initialization using a snapshot attitude estimator. This scheme also allows
initialization of the estimator with loose attitude knowledge, as may be
provided, for example, by a magnetometer whose heading measurement may be
uncertain in the presence of nearby buildings and vehicles.


\section{Tightly-Coupled Navigation Estimator}
\label{section:filter}
This section presents the remaining development of the full tightly-coupled
multi-antenna CDGNSS recursive estimator with vehicle dynamics constraints and
false integer fix mitigation.
\subsection{Propagation step}
State propagation is based on a model replacement approach in which IMU
measurements supplant a vehicle dynamics and kinematics model, permitting
broader application.

\subsubsection{Inertial measurement model}
The estimator ingests a vector of inertial measurements
$\vb u_k = \left[{\vecs f u k}{}^\tr, {\vecs {\tilde{\omega}} u k}{}^\tr
\right]^\tr$ at each epoch $k$, triggering a propagation step (time
update). The accelerometer's specific force measurement $\vecs f u k$ and the
gyroscope's angular velocity measurement $\vecs {\tilde{\omega}} u k$ are
modeled as
\begin{equation}
  \label{eq:imu_model}
\begin{aligned}
\vecs f u k = {}& \R u b \Rs b w k (\vecs a w k - \vec g w) + \vecs b u {\mathrm ak} + \vecs v u {\mathrm ak}  \\
\vecs {\tilde{\omega}} u k = {}& \vecs \omega u k + \Rs b w k \vecs \omega w \Earth + \vecs b u {\mathrm gk} + \vecs v u {\mathrm gk}
\end{aligned}
\end{equation}
where $\vecs a w k$ and $\vecs \omega u k$ are the true linear acceleration and
angular rate, respectively, of the inertial sensor, $\vec g w$ is the local
gravity vector after compensation for centripetal force due to earth rotation,
$\vecs \omega w \Earth$ is the earth's angular rate, $\vecs b u {\mathrm ak}$
and $\vecs b u {\mathrm gk}$ are time-varying zero-mean accelerometer and
gyroscope measurement biases, and $\vecs v u {\mathrm ak}$ and
$\vecs v u {\mathrm gk}$ are accelerometer and gyroscope measurement noise,
which are modeled as zero-mean white Gaussian noise.  Static biases and scale
factor errors, although not shown in (\ref{eq:imu_model}), are also modeled and
calibrated, as detailed in \cite{yoder2020visonFusion}.

\subsubsection{State dynamics}
The estimator's dynamics function
$\vb x_{k+1} = \vb f(\vb x_k, \vb u_k, \vb v_k)$ is defined by discretizing the
state dynamics, assuming a zero-order hold for the IMU measurement vector
$\vb u_k$. The accelerometer and gyroscope biases are modeled as
Ornstein-Uhlenbeck random processes with time constants $\tau_{\mr a}$ and
$\tau_{\mr g}$ and steady-state uncertainties $\sigma_{b_{\mr a}}$ and
$\sigma_{b_{\mr g}}$, respectively, as derived from IMU datasheet values (with
additional hand tuning, as datasheet values are often optimistic).  The
continuous-time state dynamics are
\begin{equation*}
\begin{aligned}
\vec {\dot r} w = {}& \vec v w \\
\vec {\dot v} w = {}& \vec a w = \R w b \R b u \left(\vecs f u k - \vecs b u {\mr a} - \vecs v u {\mr a} \right) \\
\rot {\dot R} w b = {}& \R w b \cpe{\vec \omega b} = \R w b \R b u \cpe{\vec \omega u} \\
\vecs {\dot b} u {\mr a} = {}&  -\tfrac{1}{\tau_{\mr a}} \vecs b u {\mr a} + \vecs v u {{\mr a}_2}\\
\vecs {\dot b} u {\mr g} = {}&  -\tfrac{1}{\tau_{\mr g}} \vecs b u {\mr g} + \vecs v u {{\mr g}_2}\\
\end{aligned}
\end{equation*}
where $\vecs v u {{\mr a}_2}$ and $\vecs v u {{\mr g}_2}$ are Gaussian white
noise processes driving the evolution of the accelerometer and gyroscope
biases, $\cpe{\cdot}$ denotes the skew-symmetric ``cross-product equivalent''
matrix (i.e., $\cpe{\vb a} \vb b = \vb a \times \vb b$), and where
$\vec \omega u$ is obtained by solving for it in (\ref{eq:imu_model}), which
makes it a function of the measurement $\vecs {\tilde{\omega}} u k$.  The full
discrete-time process noise vector $\vb v_k \in \mathbb{R}^{N_v}$, with
$N_v = 12$, is
\begin{align*}
\vb v_k = \left[\vecs v u {\mr a k}{}^\tr,~
\vecs v u {\mr g k}{}^\tr,~
\vecs v u {{\mr a}_2k}{}^\tr,~
\vecs v u {{\mr g}_2k}{}^\tr
\right]^{\tr}
\sim
\mathcal{N}\left(0, \vb Q_k\right)
\end{align*}
where
$$
	\vb Q_k = \mathrm{diag}\left[
		\sigma_{v_{\mr a}}^2 \vb I_{3\times 3},\;
		\sigma_{v_{\mr g}}^2 \vb I_{3\times 3},\;
		\sigma_{v_{\mr a2}}^2 \vb I_{3\times 3},\;
		\sigma_{v_{\mr g2}}^2 \vb I_{3\times 3}
	\right]
$$
The process noise parameters are determined by
$$
	\sigma_{v_{\mr a}} = \frac{S_{\mr a}}{\Delta t}, \ 
	\sigma_{v_{\mr g}} = \frac{S_{\mr g}}{\Delta t} 
$$
$$
	\sigma_{v_{\mr a 2}} = \sigma_{b_{\mr a}}^2\left(1-\exp\left(\frac{-\Delta t}{\tau_{\mr a}}\right)\right), \ 
	\sigma_{v_{\mr g 2}} = \sigma_{b_{\mr g}}^2\left(1-\exp\left(\frac{-\Delta t}{\tau_{\mr g}}\right)\right)
$$
where $S_{\mr a}$ and $S_{\mr g}$ are the accelerometer and gyroscope white noise density, respectively, and $\Delta t$ is the inertial sensor sample period. 
\subsubsection{Unscented Kalman filter propagation}
When a new inertial measurement $\vb u_k$ arrives at epoch $k$, the estimator
state is propagated from epoch $k$ to $k+1$ using the standard on-manifold UKF
propagation step.  Let $n = N_x + N_v$; then
\begin{align*}
\vb S_k = 
\chol{
	\begin{bmatrix}
	\vbh P_k & \vb 0 \\
	\vb 0 & \vb Q_k
	\end{bmatrix}
}^\tr = 
\begin{bmatrix}
\vb s_1, \vb s_2, \dots, \vb s_{n}
\end{bmatrix}
\end{align*}
A set of sigma points $\vb \chi^{(0)} \cdots \vb \chi^{(n)}$ is formed by
\begin{align*}
\vb \chi^{(0)} = {}&{}
\begin{bmatrix} \vbh x_k^\tr, & \vbh v_k^\tr \end{bmatrix}^\tr = 
\begin{bmatrix} \vbh x_k^\tr, & \vb 0^\tr 	\end{bmatrix}^\tr \\
\vb \chi^{(i)} = {}&{}
\begin{cases}
\vb \chi^{(0)} \oplus  \sqrt{n+\lambda} \,\vb s_i &\;\; i \in \{1, n\} \\
\vb \chi^{(0)} \oplus -\sqrt{n-\lambda} \,\vb s_i &\;\; i \in \{n+1,2n\}
\end{cases}
\end{align*}
The sigma points are transformed through the dynamics function
$\vb x_{k+1} = \vb f(\vb x_k, \vb u_k, \vb v_k)$ as
\begin{align*}
\vbb x_{k+1}^{(i)} = \vb f(\vb \chi^{(i)}_{\vb x}, \vb u_k, \vb \chi^{(i)}_{\vb
  v}),~~~i = 0,\dots,2n
\end{align*}
where $\vb \chi^{(i)}_{\vb x}$ and $\vb \chi^{(i)}_{\vb v}$ denote the
selection of the state and process noise components, respectively, of
$\vb \chi^{(i)}$. Finally, the sigma points are recombined to produce the
\emph{a priori} state estimate $\vbb x_{k+1}$ and covariance $\vbb P_{k+1}$:
\begin{align*}
\vbb x_{k+1} = {}& \vbh x_k \oplus \left[\sum_{i=0}^{n} w_m^{(i)}
\left(\vbb x_{k+1}^{(i)} \ominus \vbh x_{k}\right)
\right] \\
\vbb P_{k+1} = {}&\sum_{i=0}^{n} w_c^{(i)}
\left(\vbb x_{k+1}^{(i)} \ominus \vbb x_{k+1} \right)
\left(\vbb x_{k+1}^{(i)} \ominus \vbb x_{k+1} \right)^\tr
\end{align*} 
The sigma point spread parameter $\lambda$ and weights $w_m^{(i)}$, $w_c^{(i)}$
are formed as in Algorithm \ref{algo:ukf}.

\subsection{Vehicle dynamics constraints}
This \paperOrThesis{} adopts the VDC scheme of
\cite{narula2021radarpositioningjournal} with minor modifications to the NHC
sideslip model and ZUPT detection mechanism. It is described in this subsection
for completeness and notational consistency with the rest of the
\paperOrThesis{}.

\subsubsection{Non-holonomic constraint (NHC)}
\label{section:nhc}
The estimator exploits the natural constraints on the motion of four-wheeled
ground vehicles, known as non-holonomic constraints (NHCs), by casting them as
pseudo-measurements. The NHC model assumes that the vehicle rotates about a
fixed center of rotation (the origin of the \frame v frame) when a steering
input is applied, only moves in the \frame v frame $x$-direction when no
steering input is applied, does not leave the surface of the road, and
experiences only a small, predictable amount of sideslip. The vehicle sideslip
($y$-component of the velocity of the \frame v frame relative to the \frame w
frame, expressed in the \frame v frame) is modeled as dependent on the steering
rate according to the second-order polynomial model
\begin{align*}
  {\vech v v}_{k(y)} = P_0 \vecs \omega b {k(z)} + P_1
  {\left(\vecs \omega b {k(z)}\right)}^2
\end{align*}
where $[\cdot]_{(y)}$ and $[\cdot]_{(z)}$ denote taking the $y$ and $z$
components, respectively, of a vector.  The vehicle-specific polynomial
coefficients $P_0$ and $P_1$ are found via offline calibration. The estimator
adopts the following measurement model for the NHC pseudo-measurements:
\begin{align*}
\vb z_{\mr{nhc},k} = {}&{} \vb h_{\mr{nhc}}(\vb x_k) + \vb \epsilon_{\mr nhc} \in \mathbb{R}^2 \\
= {}&{} \vecs v v {(y,z)} + \vb \epsilon_{\mr {nhc}} \\
= {}&{} \left[
\R v b \left( 
\R b w \vec v w + \vec \omega b \times \left(
\vecs r b {\frame v} - \vecs r b {\frame u}
\right)
\right)
\right]_{(y,z)} + \vb \epsilon_{\mr {nhc}} \\
\vb \epsilon_{\mr {nhc}} {}&{} \sim \mathcal{N}\left(\vb 0, \begin{bmatrix}
\sigma^2_{\mr{nhc},y} & 0 \\
0 & \sigma^2_{\mr{nhc},z}
\end{bmatrix} \right)
\end{align*}
The sideslip error standard deviation $\sigma_{\mr{nhc},y}$ is
typically taken to be looser than the vertical motion error standard deviation
$\sigma_{\mr{nhc},z}$. The pseudo-measurement
\begin{align*}
\vb z_{\mr {nhc},k} = \begin{bmatrix}
{\vech v v}_{k(y)} \\ 0
\end{bmatrix}
\end{align*}
is periodically applied as an unscented measurement update to the navigation
estimator. The $\frame v$ frame origin $\vecs r b {\frame v}$ and orientation
$\R v b$ are calibrated offline in batch mode using the method developed in
\cite{narula2021radarpositioningjournal}.
\subsubsection{Zero velocity update (ZUPT)}
\label{section:zupt}
Zero-velocity updates also offer a valuable constraint on inertial sensor
drift.  Importantly for urban CDGNSS, they act as an anchor in the presence of
the worsened multipath errors that occur when the vehicle is
stopped. (Wavelength-scale vehicle motion decorrelates multipath-induced
measurement errors \citep{pesyna2016motion}.)  ZUPT pseudo-measurements are
modeled similarly to NHC pseudo-measurements but act on the full 3-dimensional
\frame v-frame velocity vector:
\begin{align*}
\vb z_{\mr {zupt},k} {}&{} = \vb h_{\mr {zupt}}\left(\vb x\right) + \vb \epsilon_{\mr {zupt}} \in \mathbb{R}^3 \\
{}&{} = \vec v v + \vb \epsilon_{\mr {zupt}}\\
{}&{} = \left[
\R v b \left( 
\R b w \vec v w + \vec \omega b \times \left(
\vecs r b {\frame v} - \vecs r b {\frame u}
\right)
\right)
\right] + \vb \epsilon_{\mr {zupt}}\\
\vb \epsilon_{\mr {zupt}} {}&{} \sim \mathcal{N}\left(\vb 0,
                              \mathrm{diag}\left[\sigma^2_{\mr {z} {x}}, \sigma^2_{\mr{z}
                              y}, \sigma^2_{\mr{z} z}\right]
\right)
\end{align*}
where $\vecs r b {\frame v}$ and $\vecs r b {\frame u}$ are the \frame b frame
positions of the \frame v and \frame u frame origins, respectively, and
$\vec \omega b = \R b u \vec \omega u$. A ZUPT pseudo-measurement
$\vb z_{\mr {zupt},k} = \vb 0$ is periodically applied when the vehicle is
detected to be stationary due to a lack of road vibration measured by the
IMU. The ZUPT measurement constraint is taken to be tighter in the $y$ and $z$
(lateral and vertical) directions, as the vehicle may appear to be stationary
via IMU vibrations when it is in reality traveling very slowly forwards.

Vehicle stationarity is detected via simple thresholding of time differences
$\Delta \vecs f u k$ and $\Delta \vecs {\tilde\omega} u k$ of raw accelerometer
and gyroscope samples, which are defined as
\begin{align*}
{\Delta \vecs f u k} \define {\vecs f u k - \vecs f u {k-1}}, \quad
{\Delta \vecs {\tilde\omega} u k} \define {\vecs {\tilde\omega} u k - \vecs {\tilde\omega} u {k-1}} 
\end{align*}
and are evaluated using threshold parameters $\gamma_{\mr{zupt,a}}$ and
$\gamma_{\mr{zupt,g}}$.  If the conditions
$\norm{\Delta \vecs f u k} < \gamma_{\mr{zupt,a}}$ and
$\norm{\Delta \vecs {\tilde\omega} u k} < \gamma_{\mr{zupt,g}}$ have both been
satisfied for all of the most recent $N_{\mr{zupt}}$ inertial measurements,
then a ZUPT pseudo-measurement is applied at every GNSS measurement
epoch. Finally, a simple $\chi^2$ innovations test is applied with threshold
parameter $P_{f,\mr{zupt}}$ to mitigate the effect of falsely-detected ZUPTs.

\subsection{Outlier Rejection using Pseudorange Innovations}
\label{section:pr_outlier}
Let
$\vb \nu_{\rho k} = \left[ \nu_{\rho k 1}, \dots, \nu_{\rho k n} \right]^\tr$
contain the elements of the GNSS innovation vector $\vb \nu_{\rm gk}$ that
correspond to pseudorange measurements. Its covariance matrix
$\vb \nu_{\rho k}$ is
\begin{align}
\label{eq:pseudorange_innovation_covariance}
\vbb P_{\rho\rho k} = 
\begin{bmatrix}
\vb  G_{1k} & \vb 0\\
\vb 0   & \vb G_{2k}
\end{bmatrix}
\vb P_{bbk}
\begin{bmatrix}
\vb G_{1k} & \vb 0\\
\vb 0   & \vb G_{2k}
\end{bmatrix}^\tr 
+ \vb \Sigma_{\rho k}
\end{align} 
where $\vb P_{bbk}$ is formed as in (\ref{eq:pbb_pbx}), and
$\vb \Sigma_{\rho k}$ is formed by selecting the rows and columns of
$\vb \Sigma_{\mr g k}$ that correspond to pseudorange measurements.  If the
estimator is consistent and the pseudorange measurements are not corrupted by
large multipath errors, then
$\nu_{\rho kn} \sim \mathcal{N}\left(0, \left(\vbb P_{\rho\rho
      k}\right)_{nn}\right)$. A test can be used to detect outlier pseudorange
measurements using the detection statistic
$$q_{kn} = \nu^2_{\rho k,n}/\left(\vbb P_{\rho\rho k}\right)_{nn}$$
with a threshold $\gamma^2$ selected to yield a sufficiently low false-positive
rate.  If a pseudorange outlier is detected for some index $n$ (i.e.,
$q_{kn} > \gamma^2$), then the corresponding DD pseudorange measurement is
assumed to be corrupted by multipath. Because the effect of multipath on GNSS
signals is primarily a function of the line-of-sight vector to the transmitter,
all pseudorange and carrier phase measurements associated with the offending
non-pivot satellite on all frequencies and baselines are assumed to be
corrupt. They are removed from the measurement vector $\vb z_{\mr g k}$ before
proceeding with the measurement update.

\subsection{False fix detection and recovery}
\label{section:false_fix}
\subsubsection{Single-epoch ambiguity resolution}
An optimal CDGNSS filter (in the maximum \emph{a posteriori} sense) must append
a new carrier phase integer ambiguity to its state each time a cycle slip is
detected in a carrier tracking loop \citep{psiaki2010kalman}.  This causes the
update and ILS solve operations to quickly become computationally intractable
when applied in an urban environment where cycle slips are common and detection
of discrete cycle slips is often impossible.

A scheme could be imagined whereby cycle slips are modeled to occur with some
probability $P_{\mr{CS}}$ at each epoch, and a suboptimal dynamic
multiple-model estimator such as the interacting multiple-model or generalized
pseudo-Bayesian estimator \citep[Sec. 11.6]{y_barshalom01_tan} is used to handle
the resulting multiple hypotheses over past cycle slips. However, for large
$P_{\mr{CS}}$, as holds for urban CDGNSS, these methods would again require
measurement updates and ILS solve operations for a quickly-growing number of
integer ambiguity candidates, making such a method computationally prohibitive
without aggressive hypothesis pruning.

This paper's estimator adopts a posture of maximum pessimism regarding cycle
slips: each carrier tracking loop in the GNSS receiver is assumed to slip
cycles between each pair of measurement epochs (i.e., $P_{\mr{CS}}=1$).
Ambiguity state growth is curtailed by discarding all ambiguity states at every
GNSS measurement epoch, either by \emph{conditioning} the state vector on the
candidate fixed solution (if the candidate fix is validated), as in
(\ref{eq:ils_solve}) and (\ref{eq:accept_fix}), or by accepting the float
solution and \emph{marginalizing} over the ambiguities, as in
(\ref{eq:accept_float}).  This single-epoch ambiguity resolution scheme is
suboptimal because it discards what continuity may be present in the integer
ambiguities from epoch to epoch, weakening the integer model strength.  But it
renders the navigation filter entirely insensitive to cycle slips, which may be
a practical necessity for urban CDGNSS, and is computationally efficient.
Moreover, the conditioning operation, when applied, greatly increases the
integer model strength for subsequent epochs, allowing the filter to ``hold
on'' to fixes by virtue of the tight epoch-to-epoch position domain constraint
provided by the inertial sensor and the vehicle dynamics pseudo-measurements.

\subsubsection{False integer fixes}
Even in the absence of modeling errors and measurement outliers, the integer
fixing procedure of Section \ref{section:sr} occasionally yields an incorrect
integer vector $\vbc n_k$ that passes ambiguity validation tests, a situation
referred to as a \emph{false integer fix} or an \emph{ambiguity resolution
  failure} \citep{green2018iagiab}. This occurs because validation tests can
only provide probabilistic guarantees of integer correctness
\citep{teunissen2009ratio}. In an urban environment, measurement outliers due to
multipath and diffraction cause the true false integer fix rate $P_f$ to
greatly exceed the specified rate $\bar P_f$ for the validation test
\citep{humphreys2019deepUrbanIts}.  Thus, conditioning the filter state on fixed
and validated integers at each epoch, as is done under a single-epoch ambiguity
resolution scheme, is perilous because false integer fixes eventually corrupt
the filter state with incorrect but highly confident priors. This causes future
GNSS measurement updates to accept a similarly incorrect fix with high
probability, repeatedly conditioning the filter state on incorrect
ambiguities. While the simple rectilinear motion model of the unaided estimator
in \cite{humphreys2019deepUrbanIts} contains sufficient process noise that the
filter eventually re-fixes to the correct ambiguities, the tight epoch-to-epoch
constraints of the present paper's tightly-coupled estimator can cause these
cycles of false fixes to persist indefinitely.

To mitigate the effect of conditioning on incorrect integer ambiguities, this
paper's estimator employs a fault detection and exclusion technique based on
solution separation. A \emph{float-only} filter, configured to never attempt
fixing integer ambiguities, is operated in parallel to the primary navigation
filter. Under the single-epoch ambiguity resolution scheme, the float-only
filter's behavior is equivalent to accepting only pseudorange measurements,
discarding carrier phase measurements entirely.

\subsubsection{Carrier phase innovations testing}
The primary filter's carrier phase measurement innovations sequence is
monitored to detect filter inconsistency, which is assumed to be caused by
false integer fixes. The carrier phase measurement NIS is defined as
$$\epsilon_{\phi k} \equiv \vbc \nu_{\phi k}^\tr \vbb P_{\phi\phi k}^{-1} \vbc
\nu_{\phi k}$$ where $\vbc \nu_{\phi k}$ is the vector of integer-resolved DD
carrier phase measurement innovations at epoch $k$, and $\vbb P_{\phi\phi k}$
is the innovation covariance matrix, formed similarly to $\vbb P_{\rho\rho k}$
of (\ref{eq:pseudorange_innovation_covariance}). The NIS $\epsilon_{\phi k}$
can be calculated during the square-root measurement update by
$\epsilon_{\phi k} = J_{2k}(\vbc n_k)$ from (\ref{eq:sr_cost}).

The test statistic used to detect false fixes is the \emph{windowed carrier
  phase NIS} $\Psi_k$ over a moving window of fixed length $l$ of past GNSS
measurement epochs. It has $N_{\Psi_k}$ degrees of freedom and is calculated by
\begin{align*}
\Psi_k      \equiv \sum_{n=k-l+1}^{k} \epsilon_{\phi n}, \quad
N_{\Psi_k}  \equiv \sum_{n=k-l+1}^{k} N_{n}
\end{align*}
where $N_k$ is the number of DD carrier phase measurements at epoch $k$. If the
filter is consistent and the integer ambiguities are correctly resolved, then
the innovations sequence should be approximately white and Gaussian, and
$\Psi_k$ should be approximately $\chi^2$-distributed with with $N_{\Psi_k}$
degrees of freedom. (This distribution is approximate due to the ``tail
clipping'' effect of integer fixing: large phase residuals are not possible
because of integer-cycle phase wrapping.)  A statistical consistency test can
be performed by choosing a desired false-alarm rate $\bar P_{f,\Psi}$ and
declaring a false fix if $\Psi_k > \gamma_{\Psi k}$, where the threshold
$\gamma_{\Psi k}$ is calculated by evaluating the inverse cumulative
distribution function (CDF) of $\chi^2(N_{\Psi_k})$ at $\bar P_{f,\Psi}$.

Use of a window of NIS values over multiple epochs increases the statistical
power of the consistency test and helps avoid premature declaration of a false
fix due to sporadic measurement outliers. However, increasing the window length
$l$ also increases the latency to detect a false fix event.

\subsubsection{False fix recovery}
\begin{figure}[h]
	\ifthesis
\else\fontsize{7}{7}\selectfont
	\fi
	\centering
	\def\svgwidth{\ifthesis\else\columnwidth\fi}
	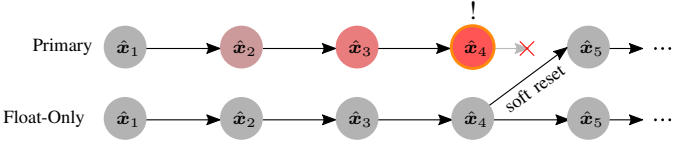
	\caption{A false-fix detection and recovery event. ``!'' denotes a
          window NIS test failure at $k=4$. The primary navigation filter's
          state estimate and covariance matrix are assumed to be contaminated
          by false integer fixes at past epochs and are discarded. They are
          replaced after epoch $4$ with the less-certain but uncontaminated
          state and covariance of the float-only filter.}
	\label{fig:soft_reset}
      \end{figure}

If a false fix is detected ($\Psi_k > \gamma_{\Psi k}$), the estimator performs
a \emph{soft reset}, discarding the primary navigation filter's state estimate
and covariance and replacing them with a copy of the float-only filter's state
and covariance, as shown in Fig. \ref{fig:soft_reset}.
      \begin{figure}[h]
	\ifthesis
\else\fontsize{7}{7}\selectfont
	\fi
	\centering
	\def\svgwidth{\ifthesis\else\columnwidth\fi}
	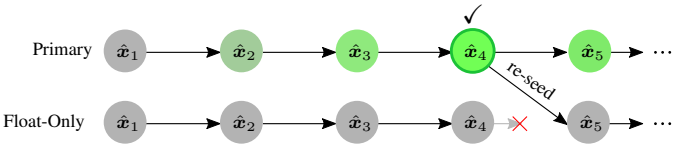
	\caption{A ``re-seed'' event.``\checkmark'' denotes re-seed criteria
          being met at epoch $4$, indicating very high confidence in the
          correctness of the primary filter's integer fix and consistency of
          its state estimate. The estimator replaces the state estimate and
          covariance of the float-only filter with a copy of that of the
          primary navigation filter.}
	\label{fig:reseed}
\end{figure}


\subsubsection{Float-only estimator re-seeding}
\label{section:reseed}
To increase the probability of a correct fix after a soft reset, the
``float-only'' filter is occasionally \emph{re-seeded} with the state of the
primary filter during epochs over which a set of heuristic criteria indicate
that a correct fix is extremely likely. This operation carries the risk that
the float-only filter could also be contaminated with information from a false
integer fix in the primary filter. But in practice, heuristic criteria can be
set to strictly limit this event's probability.  The four criteria used in the
next section's evaluation are given in Table \ref{table:reseed_criteria}.

\begin{table}[h]
\centering
\def\arraystretch{1.4}
{
\ifthesis
\else
	\ifwiley
	\fontsize{8}{8}\selectfont
	\fi
\fi
\begin{tabular}{r r l} \toprule
  Carrier phase measurement NIS & $\sfrac{\epsilon_{\phi k}}{N_{k}}$ & $\leq 1.0$ \\
  Windowed carrier phase NIS & $\sfrac{\Psi_k}{N_{\Psi k}}$ & $\leq 0.5$ \\
  Last fix number of DD measurements & $N_{k}$ & $\geq 10$ \\
  Time since last soft reset & $t_{\mr{sr},k}$ & $\geq 2.0$ s \\
  \bottomrule
\end{tabular}
}
\caption{Re-seed criteria used in the evaluation of Section
  \ref{section:results}. A re-seed operation is performed if all four
  conditions are met.}
\label{table:reseed_criteria}
\end{table}
One might argue that these criteria for re-seeding the float-only filter are
redundant because an integer aperture test for validating the primary filter's
integer estimate can be made arbitrarily strict, obviating additional
validation. But integer aperture theory is founded on modeling measurement
error distributions as Gaussian \citep{teunissen2003integer}, which is a poor
approximation in the urban environment, leading to low fixed solution
availability \citep{humphreys2019deepUrbanIts}.  Teunissen's recent extension of
so-called best integer equivariant estimation to the class of elliptically
contoured distributions in \cite{teunissen2020best} may offer a means of
providing a better re-seed estimate for the float-only filter in urban
environments, but it has not been tested with empirical urban data.  Meanwhile,
application of this paper's re-seeding technique with the criteria in Table
\ref{table:reseed_criteria} will be shown in the next section to significantly
increase integer fix availability while respecting a low false fix rate.



\newcommand\tsection[1]{
	\ifthesis\else\section{#1}\fi
}

\newcommand\tsubsection[1]{
	\ifthesis\else\subsection{#1}\fi
}

\newcommand\tsubsubsection[1]{
	\ifthesis\else\subsubsection{#1}\fi
}

\tsection{Performance Evaluation in a Deep Urban Environment}
\label{section:results}

\begin{figure}[h]
  \centering
  \includegraphics[width={\ifthesis\else{200pt}\fi}]{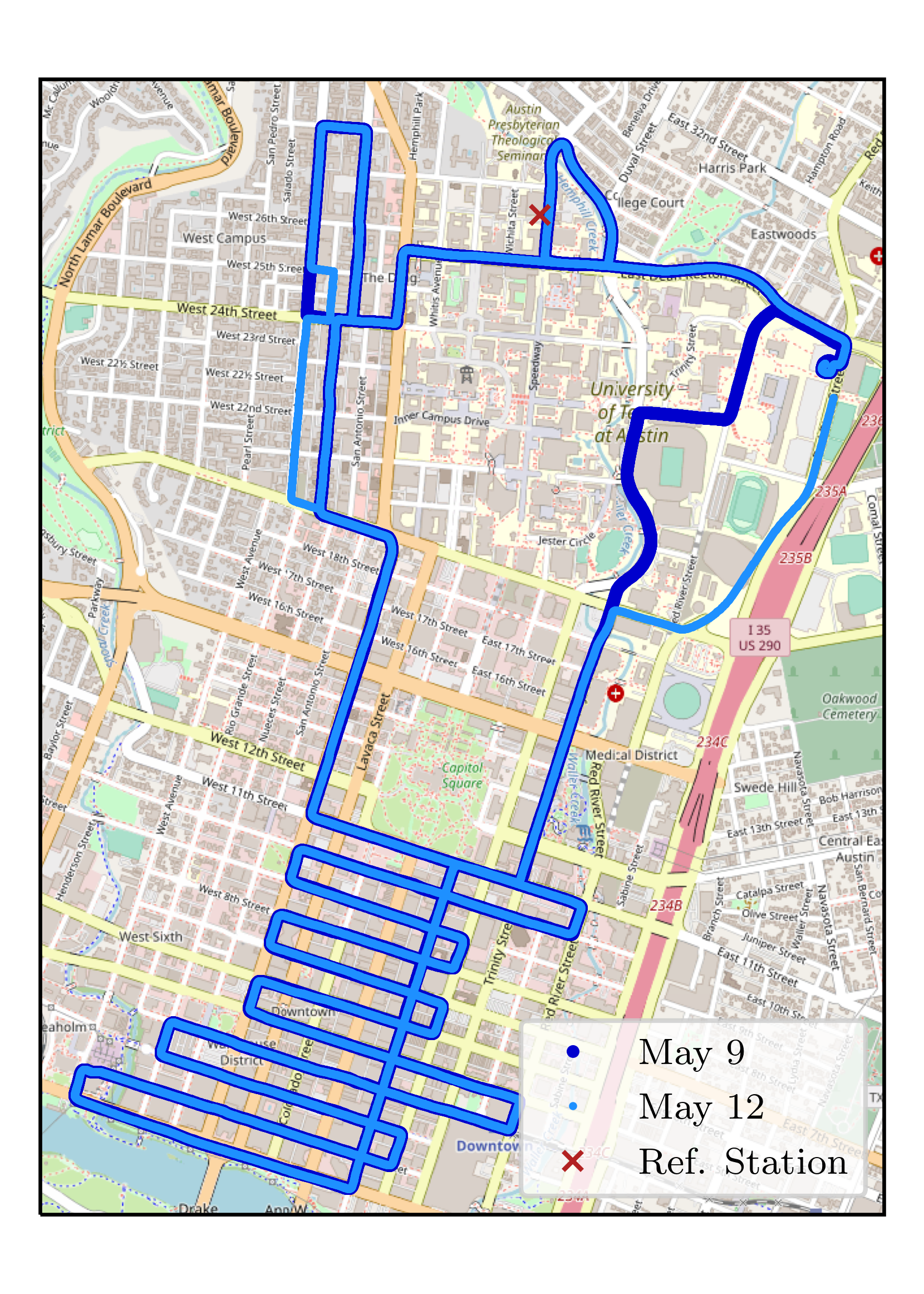}
  \caption{Overview of the CDGNSS reference station position and routes
    driven through the urban core of Austin, Texas in the TEX-CUP urban
    positioning datasets. Routes differ slightly from May 9 to May 12
    due to road closures.}
  \label{fig:map}
\end{figure}

\tsubsection{Experimental setup}
\label{section:setup}
The tightly-coupled CDGNSS estimator described in the foregoing sections was
implemented in C++ as a new version of the \ppe sensor fusion engine
\citep{humphreys2019deepUrbanIts}, and was experimentally evaluated against the
publicly-available TEX-CUP urban positioning dataset.  TEX-CUP comprises raw
GNSS intermediate-frequency (IF) samples and inertial data collected on 9 and
12 May, 2019, using the University of Texas \emph{Sensorium} vehicular
perception research platform \citep{narula2020texcup}. The dataset consists of a
total of over 2 hours of driving in Austin, Texas in conditions ranging from
light to dense urban; routes are shown in Fig. \ref{fig:map}.
      
Two-bit-quantized IF samples were captured at the Sensorium and at the
reference station through \emph{RadioLynx}, a low-cost L1+L2 GNSS front end
with a 5 MHz bandwidth at each frequency, and were processed with the \pprx
software-defined GNSS receiver \citep{humphreys2019deepUrbanIts}. The Sensorium
RadioLynx was connected to two Antcom G8 GNSS antennas separated by 1.0668
meters in the vehicle Y direction, and the reference RadioLynx was connected to
a Trimble Zephyr II geodetic-grade GNSS antenna.

The system's performance was separately evaluated using inertial data from each
of the Sensorium's two MEMS inertial sensors. The first, a LORD MicroStrain
3DM-GX5-25, is an industrial-grade sensor. The second, a Bosch BMX055, is a
surface-mount consumer-grade sensor. Their relevant datasheet specifications
are compared in \ifthesis
\else
\ifwiley{}\else{reference}\fi{} \cite{yoder2021thesis}.
\fi

The system's positioning performance was evaluated by comparing against
TEX-CUP's forward-backward smoothed ground-truth reference trajectory. This
ground truth was generated by post-processing the data from an iXblue ATLANS-C
mobile mapping system comprising a Septentrio AsteRx4 RTK receiver and a
high-end tactical-grade IMU. The reported accuracy of the ground truth
trajectory varied between 2 and 15 cm (1-$\sigma$) along the route. Because it
is impossible to directly evaluate integer ambiguity resolution performance,
the integer fixing performance in the following sections was evaluated by
considering integer fixes to be correct if the 3D distance to the ground truth
was below 30 cm, following \cite{humphreys2019deepUrbanIts}.

Because TEX-CUP contains several minutes of no motion in an open-sky
environment at the beginning and end of each capture, the estimator was run on
a subset of each capture beginning approximately 10 seconds before first
motion and ending 10 seconds after last motion. On the May 9 dataset, the
estimator was run and evaluated from GPS time of week (TOW) 411003 s to 415029
s, and on the May 12 dataset, from TOW 63770 s to 67972 s.

\tsubsection{Baseline configuration}
\label{section:config}
The performance of the tightly-coupled estimator with all proposed features and
signals enabled (the ``baseline configuration'') was evaluated using each IMU
on the May 9 and May 12 TEX-CUP datasets. \pprx was configured as described in
\cite{humphreys2019deepUrbanIts}, tracking the GPS L1 C/A, GPS L2C (combined
CL+CM codes), Galileo E1 (combined B+C codes), and L1 SBAS (WAAS) signals on
the reference and both rover antennas. Data bit prediction and wipeoff were
performed on the GPS L1 C/A and SBAS signals. Reference and rover GNSS
observables were produced at a rate of 5 Hz. \ppe's baseline configuration
parameters are given in Table \ref{table:ppengine_config}.

\begin{table}[t]
	\centering
	\def\arraystretch{1.0}
	\setlength{\tabcolsep}{2pt}
	{
		\ifthesis
\else
			\ifwiley
				\fontsize{8}{8}\selectfont
			\fi
		\fi
	\begin{tabular}{r r l} \toprule
		\\[-9pt] 
		\multicolumn{3}{l}{\textbf{CDGNSS parameters}}\\
		Carrier-to-noise ratio threshold & $\sfrac{C}{N_0}$&$\geq 40$ dB-Hz \\
		Phase lock statistic threshold & $s_{\theta}$&$\geq 0.8$ \\
		Elevation mask & $\theta_{\mr{el}}$&$\geq 10^\circ$ \\
		Integer aperture (IA) validation test  & \multicolumn{2}{c}{FF-difference test \ifwiley{}\else{\cite{wang2015iafit}}\fi{}} \\
		IA fixed failure rate & $\bar P_f$&$= 0.001$ \\
		Undifferenced zenith pseudorange std & $\sigma_{\rho}$&$= 1.5$ m \\
		Undifferenced zenith phase std & $\sigma_{\phi}$&$= 0.006$ m \\
		Pseudorange outlier threshold std & $\gamma$&$=1.5\sigma$ \\
		False fix detection window length & $l$&$= 10$ \\
		False fix detection threshold & $\bar P_{f,\Psi}$&$= 10^{-15}$\\ \midrule
		\multicolumn{3}{l}{\textbf{IMU parameters}}\\
		Accelerometer noise density & $\sqrt{S_{\mr a}}$&$= 100, 300\  \mu g/\sqrt{\mr{Hz}}^*$\\
		Accelerometer bias steady-state std & $\sigma_{b_{\mr a}}$&$= 0.5, 10\ \mr{m}g^*$\\
		Accelerometer bias time constant & $\tau_{\mr a}$&$= 100$ s\\
		Gyroscope noise density & $\sqrt{S_{\mr g}}$&$= 0.01, 0.05\ ({}^\circ/\mr{s})/\sqrt{\mr{Hz}}^*$\\
		Gyroscope bias steady-state std & $\sigma_{b_{\mr g}}$&$= 8, 30\ {}^\circ/\mr{hr}^*$ \\
		Gyroscope bias time constant & $\tau_{\mr g}$&$= 100$ s\\ \midrule
		\multicolumn{3}{l}{\textbf{NHC parameters}}\\ 
		Lateral std & $\sigma_{\mr{nhc},y}$&$= 0.1$ m/s \\
		Vertical std & $\sigma_{\mr{nhc},z}$&$= 0.2$ m/s\\ \midrule
		\multicolumn{3}{l}{\textbf{ZUPT parameters}}\\
		Longitudinal std &$\sigma_{\mr zx}$&$ = 0.05$ m/s\\
		Lateral/vertical std &$\sigma_{\mr zy},\sigma_{\mr zz}$&$ = 0.01$ m/s\\
		Accelerometer noise threshold & $\gamma_{\mr{zupt,a}}$&$ = 0.8 $ m/s$^2$\\
		Gyroscope noise threshold & $\gamma_{\mr{zupt,g}}$&$ = 0.006, 0.018$ rad/s*\\
		ZUPT detection window & $N_{\mr{zupt}}$&$ = 10, 30$*\\
		Innovations test threshold & $\bar P_{f,{\mr{zupt}}}$&$ = 10^{-30}, 10^{-6}$*\\
		\bottomrule
	\end{tabular}
}
\caption{Baseline \ppe configuration parameters. Pairs marked with ``*''
  indicate separate parameter values used with the industrial-grade and
  consumer-grade IMU, respectively. IMU noise parameters were increased from
  datasheet values for consistency with empirical observations.}
	\label{table:ppengine_config}
\end{table}

The phase center variation of the Sensorium antennas with respect to signal
elevation angle was calibrated as in \cite{humphreys2019deepUrbanIts}. The
orientation, body-frame position, axis scale factors, and steady-state biases
of both inertial sensors were calibrated offline using a short period of
dynamic open-sky GNSS data at the beginning of the May 12 dataset.

The position and attitude states of the tightly-coupled navigation estimator
were initialized with the first available batch of GNSS observables and
inertial measurements. The position state was initialized with the standard
(single-ended) pseudorange position solution for the primary antenna. The
attitude state was initialized by combining the gravity vector as determined by
the IMU's accelerometers with a constrained-baseline snapshot CDGNSS solution
for the baseline connecting the primary and secondary vehicle antennas, as
determined with a brute-force attitude-domain search.

\tsubsection{Baseline performance}
\label{section:baseline_perf}

\ifthesis\else
	\begin{figure}[t]
	\centering
	\includegraphics[width={\ifthesis{0.75\columnwidth}\else\columnwidth\fi}]{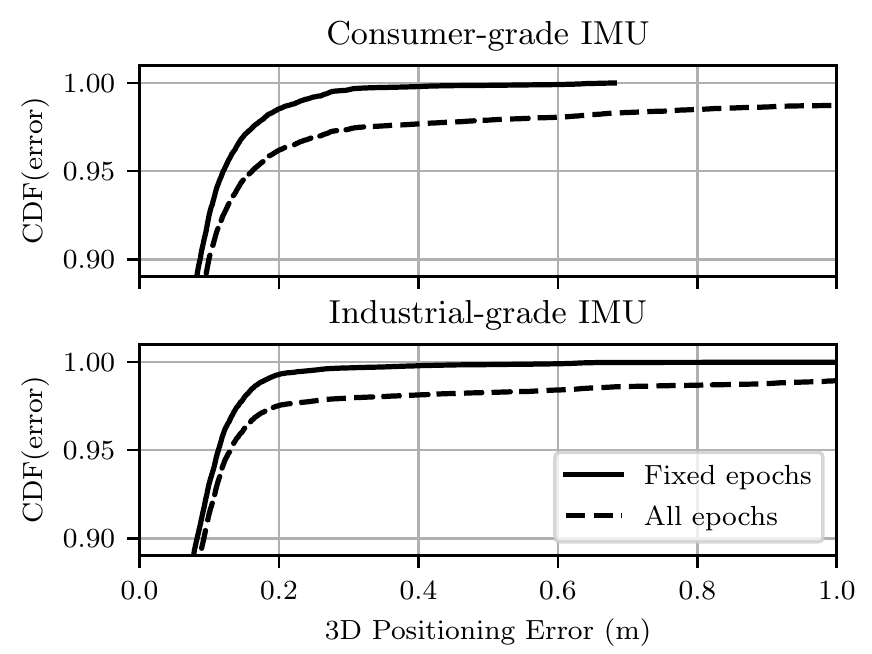}
	\caption{Cumulative distribution functions of 3D positioning error of
          baseline estimator configuration across both days of the TEX-CUP
          urban positioning dataset.}
	\label{fig:cdf}
\end{figure}

\ifthesis\begin{landscape}\fi
	\begin{table*}[h]
		\centering
		\def\arraystretch{1.2}
		\setlength\tabcolsep{6pt}
		\newcolumntype{d}[1]{S[table-format={#1}]}
		{
			\ifthesis
			\fontsize{10}{10}\selectfont
			\def\arraystretch{1.4}
			\setlength\tabcolsep{2pt}
			\else
				\ifwiley
				\fontsize{8}{8}\selectfont
				\fi
			\fi
			\begin{tabular}{r r d{2.2} d{2.2} d{4.1} d{4.1} d{3.1} d{4.1} d{3.1} d{3.1}}
				\toprule
				& & & & \multicolumn{6}{c}{Overall positioning performance (fix \& float epochs)} \\ \cmidrule(r){5-10}
				& & \multicolumn{2}{c}{Ambiguity Resolution} &  \multicolumn{2}{c}{3D}  & \multicolumn{2}{c}{Horizontal}  & \multicolumn{2}{c}{Vertical}\\ \cmidrule(r){3-4} \cmidrule(r){5-6} \cmidrule(r){7-8} \cmidrule(r){9-10}
				\multicolumn{2}{c}{Dataset} & 
				{$P_V$ (\%)} & 
				{$P_f$ (\%)} & 
				{$d_{95}$ (cm)} & 
				{RMSE (cm)} & 
				{$d_{95\mr{h}}$ (cm)} & 
				{RMSE (cm)} &
				{$d_{95\mr{v}}$ (cm)} & 
				{RMSE (cm)}  \\ \midrule		
				May 9 & Unaided & \unaidedmayninelynxPv & \unaidedmayninelynxPf & \unaidedmayninelynxpnftotal & \unaidedmayninelynxrmse & \unaidedmayninelynxpnfHorzTotal & \unaidedmayninelynxrmseHorz & \unaidedmayninelynxpnfVertTotal & \unaidedmayninelynxrmseVert \\
				& Consumer-grade IMU & \baselinemayninelynxPv & \baselinemayninelynxPf & \baselinemayninelynxpnftotal & \baselinemayninelynxrmse & \baselinemayninelynxpnfHorzTotal & \baselinemayninelynxrmseHorz & \baselinemayninelynxpnfVertTotal & \baselinemayninelynxrmseVert \\
				& Industrial-grade IMU & \baselinemayninelordPv & \baselinemayninelordPf & \baselinemayninelordpnftotal & \baselinemayninelordrmse & \baselinemayninelordpnfHorzTotal & \baselinemayninelordrmseHorz & \baselinemayninelordpnfVertTotal & \baselinemayninelordrmseVert \\ \midrule
				May 12 & Unaided & \unaidedmaytwelvelynxPv & \unaidedmaytwelvelynxPf & \unaidedmaytwelvelynxpnftotal & \unaidedmaytwelvelynxrmse & \unaidedmaytwelvelynxpnfHorzTotal & \unaidedmaytwelvelynxrmseHorz & \unaidedmaytwelvelynxpnfVertTotal & \unaidedmaytwelvelynxrmseVert \\
				& Consumer-grade IMU & \baselinemaytwelvelynxPv & \baselinemaytwelvelynxPf & \baselinemaytwelvelynxpnftotal & \baselinemaytwelvelynxrmse & \baselinemaytwelvelynxpnfHorzTotal & \baselinemaytwelvelynxrmseHorz & \baselinemaytwelvelynxpnfVertTotal & \baselinemaytwelvelynxrmseVert \\
				& Industrial-grade IMU & \baselinemaytwelvelordPv & \baselinemaytwelvelordPf & \baselinemaytwelvelordpnftotal & \baselinemaytwelvelordrmse & \baselinemaytwelvelordpnfHorzTotal & \baselinemaytwelvelordrmseHorz & \baselinemaytwelvelordpnfVertTotal & \baselinemaytwelvelordrmseVert \\
				\bottomrule
			\end{tabular}
		}
		\caption {Baseline estimator ambiguity resolution and
                  positioning performance on each day of the TEX-CUP
                  dataset. ``Unaided'' indicates the use of the motion model of
                  \cite{humphreys2019deepUrbanIts} in lieu of inertial tight
                  coupling, as described in Sec. \ref{section:unaided}. Quoted
                  95th percentile and RMS error quantities are over the entire
                  dataset (i.e., for both float and fixed epochs).  $P_V$
                  denotes the availability of an aperture-test-validated fixed
                  solution for conditioning in the primary filter.  $P_f$
                  denotes the false fix rate, as determined by an excursion of
                  the primary filter beyond 30 cm from the ground truth when
                  conditioned on fixed ambiguities.}
		\label{table:baseline}
	\end{table*}
	\ifthesis\end{landscape}\fi

\ifthesis\begin{landscape}\fi
	\begin{figure*}[h]
		\centering
		\includegraphics[width={\ifthesis{7in}\else\textwidth\fi}]{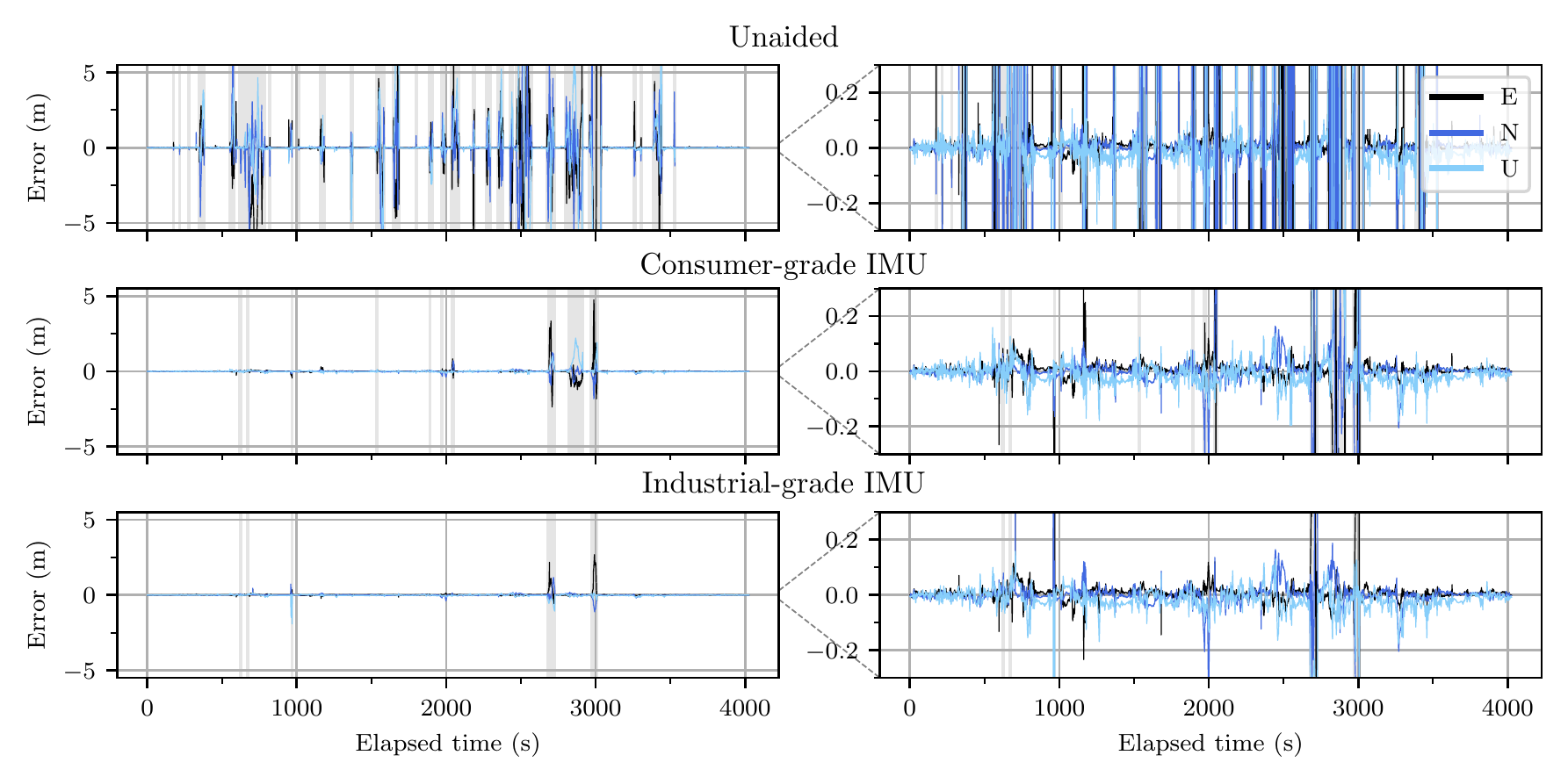}
		\caption{Baseline estimator positioning error over time in the
                  East, North, and Up directions for the May 9 TEX-CUP
                  dataset. This day had worse GNSS satellite geometry and
                  therefore lower positioning performance than on May 12. Gray
                  shading indicates float epochs (periods when the estimator
                  accepted the float CDGNSS solution). ``Elapsed time''
                  indicates time since dataset start at GPS TOW 411003 s.}
		\label{fig:poserr}
	\end{figure*}
	\ifthesis\end{landscape}\fi
	\begin{figure}[h]
	\centering
	\includegraphics[width={\ifthesis{0.75\columnwidth}\else\columnwidth\fi}]{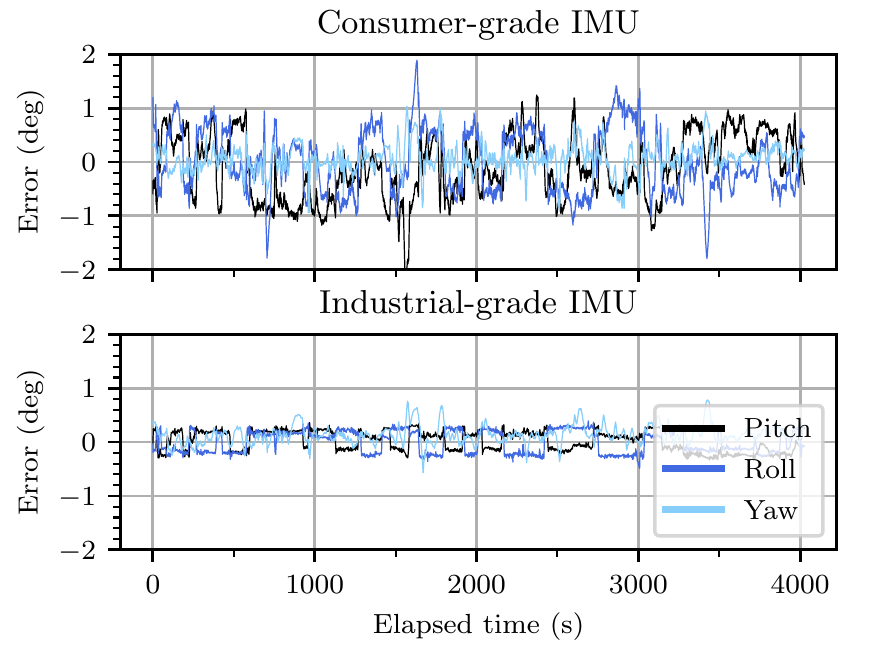}
	\caption{Baseline estimator attitude estimation error over time for the
          May 9 TEX-CUP dataset. Heading-dependent pitch and roll errors are
          evident when tightly coupled with the industrial-grade IMU; these are
          likely due to residual calibration errors.}
	\label{fig:atterr}
\end{figure}

\begin{table}[h]
	\centering
	\def\arraystretch{1.2}
	\newcolumntype{d}[1]{S[table-format={#1}]}
	{
		\ifthesis
		\fontsize{10}{10}\selectfont
		\def\arraystretch{1.4}
		\else
			\ifwiley
			\fontsize{8}{8}\selectfont
			\fi
		\fi
		\begin{tabular}{r c c c c c c}
			\toprule
			&  \multicolumn{2}{c}{Roll (\textdegree)}  & \multicolumn{2}{c}{Pitch (\textdegree)}  & \multicolumn{2}{c}{Yaw (\textdegree)}\\ \cmidrule(r){2-3} \cmidrule(r){4-5} \cmidrule(r){6-7}
			\multicolumn{1}{c}{Dataset} & 
			{$p_{95}$} & 
			{RMS} & 
			{$p_{95}$} & 
			{RMS} & 
			{$p_{95}$} & 
			{RMS}   \\
			\midrule	
			\multicolumn{1}{l}{May 9} & & & & &\\ 
			Consumer-grade IMU & \baselinemayninelynxpnfRoll & \baselinemayninelynxrmsRoll &  \baselinemayninelynxpnfPitch & \baselinemayninelynxrmsPitch & \baselinemayninelynxpnfYaw & \baselinemayninelynxrmsYaw\\
			Industrial-grade IMU & \baselinemayninelordpnfRoll & \baselinemayninelordrmsRoll &  \baselinemayninelordpnfPitch & \baselinemayninelordrmsPitch & \baselinemayninelordpnfYaw & \baselinemayninelordrmsYaw\\ \midrule
			\multicolumn{1}{l}{May 12} & & & & &\\ 
			Consumer-grade IMU & \baselinemaytwelvelynxpnfRoll & \baselinemaytwelvelynxrmsRoll &  \baselinemaytwelvelynxpnfPitch & \baselinemaytwelvelynxrmsPitch & \baselinemaytwelvelynxpnfYaw & \baselinemaytwelvelynxrmsYaw\\
			Industrial-grade IMU & \baselinemaytwelvelordpnfRoll & \baselinemaytwelvelordrmsRoll &  \baselinemaytwelvelordpnfPitch & \baselinemaytwelvelordrmsPitch & \baselinemaytwelvelordpnfYaw & \baselinemaytwelvelordrmsYaw\\ \bottomrule
			
		\end{tabular}
	}
	\caption {Baseline estimator attitude performance on each day of the
          TEX-CUP dataset. Quoted 95th percentile and RMS error quantities are
          over the entire dataset (i.e., for both float and fixed epochs).}
	\label{table:baseline_attitude}
\end{table}
\fi
\tsubsubsection{Ambiguity resolution and positioning}

The achieved integer-fix availability was \baselinelordPv{} and
\baselinelynxPv{} when tightly coupled with the industrial-grade and
consumer-grade IMUs, respectively, across both days of the dataset.  When
tightly coupled with the industrial-grade IMU, the 95th-percentile horizontal
positioning error was \baselinelordpnfHorzFix{} cm when fixed and
\baselinelordpnfHorzTotal{} cm overall (fixed and float). Using the
consumer-grade IMU, the 95th-percentile horizontal error was
\baselinelynxpnfHorzFix{} cm when fixed and \baselinelynxpnfHorzTotal{} cm
overall.  The empirical CDF of 3D positioning errors using both grades of
inertial sensor is shown in Fig. \ref{fig:cdf}.  Detailed statistics of the
estimator's positioning performance in the baseline configuration are given in
Table \ref{table:baseline}, and statistics of its attitude performance in Table
\ref{table:baseline_attitude}.  Figs. \ref{fig:poserr} and \ref{fig:atterr}
show the position and attitude error, respectively, over time for the May 9
portion of TEX-CUP.

\ifthesis\fi

\tsubsubsection{Attitude}

The attitude performance of the baseline estimator is excellent when tightly
coupled with either the industrial-grade or consumer-grade inertial sensor,
achieving single-degree-level precision in all three axes with the
consumer-grade sensor, and sub-degree precision with the inertial
sensor. Better performance with the industrial-grade sensor is as expected due
to its significantly better gyroscope noise properties.

\ifthesis\fi

\tsubsection{Effect of inertial tight coupling}
\label{section:unaided}

To evaluate the benefit of inertial tight coupling, \ppe was run in an
``unaided'' mode, using the nearly-constant-velocity motion model described in
\cite{humphreys2019deepUrbanIts} for propagation in place of an inertial
sensor. Attitude dynamics were modeled as a simple integrated white noise
process, with noise intensity of 0.3 $\sfrac{\circ}{\sqrt{\mr{s}}}$ in the
vehicle pitch and roll axes, and 5.7 $\sfrac{\circ}{\sqrt{\mr{s}}}$ in the
vehicle yaw axis. Because vehicle pitch (rotation about the axis connecting the
primary and secondary Sensorium antennas) is not strongly observable, a weak
pseudo-measurement of zero pitch was added with standard deviation $10^\circ$
at a rate of 5 Hz, which was found to provide good estimation performance.

The positioning and ambiguity resolution performance of the unaided estimator
is shown in Table \ref{table:baseline}. Tight coupling with even a
consumer-grade IMU has a clearly beneficial effect on ambiguity resolution,
greatly increasing the fraction of fixed-integer epochs for comparable $P_f$,
and reducing 95th-percentile and RMS positioning errors from meters to
centimeters.

\tsubsection{Performance in alternate configurations}
\label{section:alternate_configs}

Next, the estimator was run in a collection of alternate configurations (with
various features disabled) in order to study its performance's sensitivity to
the presence of various algorithmic components. Results are given in Table
\ref{table:degradations}.

\newcounter{scenarionum}
\newcommand{\scenario}[1]{\refstepcounter{scenarionum}\label{#1}}
\ifthesis\begin{landscape}\fi
\begin{table*}[h]
	\centering
	\def\arraystretch{1.2}
	\newcolumntype{d}[1]{S[table-format={#1}]} 
	{
	\ifthesis
\else
		\ifwiley
		\fontsize{8}{8}\selectfont
		\fi
	\fi
	\begin{tabular}{r r d{2.2}d{2.2}d{3.1}d{3.1} d{2.2}d{2.2}d{3.1}d{3.1}}
		\toprule
		& & \multicolumn{4}{c}{Industrial-grade IMU} & \multicolumn{4}{c}{Consumer-grade IMU} \\ \cmidrule(r){3-6} \cmidrule(r){7-10}
		\multicolumn{2}{c}{Configuration} & 
		{$P_V$ (\%)} & 
		{$P_f$ (\%)} & 
		{$d_{95}$ (cm)} & 
		{RMSE (cm)} & 
		{$P_V$ (\%)} & 
		{$P_f$ (\%)} & 
		{$d_{95}$ (cm)} & 
		{RMSE (cm)} \\
		\midrule
		\scenario{scenario:baseline} \ref{scenario:baseline}&Baseline (all features enabled)& 97.52 & 0.33 & 13.0 & 17.5& 96.62 & 0.38 & 16.2 & 24.8\\ 
\scenario{scenario:single-antenna} \ref{scenario:single-antenna}&Single vehicle antenna& 96.89 & 0.55 & 16.2 & 25.3& 95.73 & 0.53 & 19.3 & 30.6\\ 
\scenario{scenario:no-vvc} \ref{scenario:no-vvc}&Sans non-holonomic constraints (\S \ref{section:nhc})& 97.48 & 0.33 & 13.2 & 18.3& 94.86 & 0.60 & 26.1 & 116.0\\ 
\scenario{scenario:no-zupt} \ref{scenario:no-zupt}&Sans zero-velocity updates (\S \ref{section:zupt})& 97.39 & 0.34 & 13.8 & 38.2& 96.39 & 0.40 & 16.9 & 27.1\\ 
\scenario{scenario:no-pr-outlier} \ref{scenario:no-pr-outlier}&Sans pseudorange outlier exclusion (\S \ref{section:pr_outlier})& 93.09 & 0.48 & 19.1 & 45.5& 87.73 & 1.25 & 241.8 & 133.6\\ 
\scenario{scenario:no-reseed} \ref{scenario:no-reseed}&Sans re-seed (\S \ref{section:reseed})& 97.00 & 0.91 & 15.0 & 36.8& 96.58 & 0.46 & 16.5 & 34.2\\ 
\scenario{scenario:no-mm} \ref{scenario:no-mm}&Sans false-fix detection \& recovery (\S \ref{section:false_fix})& 99.00 & 24.28 & 590.8 & 194.7& 97.27 & 10.03 & 121.6 & 46.2\\ 
\scenario{scenario:single-freq} \ref{scenario:single-freq}&Single frequency (L1 only)& 97.65 & 0.96 & 14.7 & 24.2& 90.35 & 0.38 & 138.9 & 87.1\\ 
\scenario{scenario:no-sbas} \ref{scenario:no-sbas}&Sans SBAS& 95.73 & 0.55 & 17.1 & 25.6& 88.86 & 0.86 & 191.2 & 80.3\\ 
\scenario{scenario:ekf-update} \ref{scenario:ekf-update}&EKF CDGNSS Update (\texttt{linearizeEkf})& 97.52 & 0.33 & 13.0 & 17.5& 96.62 & 0.38 & 16.2 & 24.8\\ 

		\bottomrule
	\end{tabular}
	}
	\caption {Estimator ambiguity resolution and positioning performance on
          the combined TEX-CUP May 9 and May 12 datasets with various estimator
          features disabled.}
	\label{table:degradations}
\end{table*}
\ifthesis\end{landscape}\fi

\tsubsubsection{Performance in single-antenna mode}

In Configuration \ref{scenario:single-antenna}, the estimator was run using
only a single vehicle-mounted antenna. CDGNSS observables were formed for only
a single baseline, between the reference antenna and the primary Sensorium GNSS
antenna. Despite having fewer integer ambiguities (only one baseline's worth)
to fix at each measurement epoch, the integer fix availability was lower than
for the multi-antenna case in this configuration because the estimator was no
longer able to exploit the measurement noise cross-covariance between baselines
due to the shared antenna. Positioning performance is slightly worse in the
single-antenna case by all metrics, as would be expected due to the loss of
half of all GNSS measurements.

\tsubsubsection{Performance without vehicle motion constraints}

The non-holonomic constraints (Sec. \ref{section:nhc}) were disabled in
Configuration \ref{scenario:no-vvc}, and zero-velocity updates
(Sec. \ref{section:zupt}) were disabled in Configuration
\ref{scenario:no-zupt}.  It can be seen that the non-holonomic constraints have
only a small effect on performance when tightly coupling with the
industrial-grade IMU. With the consumer-grade IMU, however, the effect is much
greater. Clearly, non-holonomic constraints (Sec. \ref{section:nhc}) provide a
major performance boost in the consumer-grade IMU case, cutting the number of
float GNSS measurement epochs in half (fix availability increased from
\novvclynxPv{} to \baselinelynxPv{}).

The incorporation of zero-velocity updates improves all of the presented
statistics when tightly coupling with either the industrial-grade or
consumer-grade IMU. But for the consumer-grade IMU, the benefit of ZUPTs is
minimal. This is likely because the poor noise properties of the consumer-grade
IMU required such strict thresholds to limit false detections that many
opportunities to apply a ZUPT were missed. However, ZUPTs have a clearly
positive benefit on overall RMS position error, as they help to constrain
against float-solution position error when the vehicle is stopped, which is
when urban code multipath errors are largest.

\tsubsubsection{Performance without pseudorange outlier exclusion}

The pseudorange innovations-based outlier exclusion mechanism
(Sec. \ref{section:pr_outlier}) was disabled in Configuration
\ref{scenario:no-pr-outlier}. Without this mechanism, the availability of
validated integer-fixed solutions decreased drastically due to the presence of
outlier measurements caused by multipath. Interestingly, the false fix rate
$P_f$ was elevated, but not to extreme levels. This was likely because false
fixes due to multipath-induced outliers were reverted by the false-fix recovery
mechanism.

\tsubsubsection{Performance without false-fix detection and recovery}

In Configuration \ref{scenario:no-reseed}, the ``re-seed'' mechanism described
in Sec. \ref{section:reseed} was disabled, and in Configuration
\ref{scenario:no-mm} the entire false-fix detection and recovery mechanism
(Sec. \ref{section:false_fix}) was disabled. Integer fix performance without
the re-seed mechanism is appreciably reduced, as the estimation performance of
the float-only estimator suffers without the ability to re-seed from especially
trustworthy integer fixes.  Disabling the false-fix detection and recovery
mechanism has a catastrophic effect on false fixing rate $P_f$, for the reasons
given in Sec. \ref{section:false_fix}.

\tsubsubsection{Performance on subsets of GNSS signals}

In Configuration \ref{scenario:single-freq}, the estimator was run using only
L1 GNSS signals (i.e., GPS L2C was disabled), and in Configuration
\ref{scenario:no-sbas}, SBAS L1 signals were disabled. Ambiguity resolution and
positioning performance on these GNSS signal subsets was fairly close to the
baseline case when using the industrial-grade IMU, but with the consumer-grade
IMU a substantial loss of integer-fix availability occurred (down from
\baselinelynxPv{} to \singlefreqlynxPv{} and \nosbaslynxPv{} in each of these
configurations, respectively). The weaker motion constraints provided by the
consumer-grade IMU cause the estimator to require more GNSS signals for
acceptable integer fix availability.

\tsubsubsection{Performance with EKF-based linearization}

The estimator was run using \texttt{linearizeEkf} in place of
\texttt{linearizeUkf} in Configuration \ref{scenario:ekf-update}. Due to the
excellent attitude performance of the estimator with either inertial sensor, no
significant difference in performance arises on the TEX-CUP dataset by using
the UKF update. This can be explained by referring to the results of
\ifthesis
\else
Fig.
\fi
\ref{fig:linearization_montecarlo_ia}, which show that the benefit of the UKF
linearization is significant for the Sensorium's inter-antenna distance only
when attitude uncertainty exceeds approximately $8^\circ$ on a single axis,
whereas the attitude error on the TEX-CUP dataset never exceeded $2^\circ$.


\section{Concluding Remarks}
\label{section:conclusions}
A vehicular pose estimation technique has been presented and evaluated that
tightly-couples multi-antenna CDGNSS, a low-cost MEMS IMU, and vehicle dynamics
constraints (non-holonomic constraints and zero-velocity updates). The
unscented transform was used to linearize the multi-antenna CDGNSS update,
allowing the use of a linear integer least squares solver for ambiguity
resolution while exploiting between-baseline correlations and respecting the
constraints provided by known vehicle antenna geometry, even under large
attitude uncertainties. Robust estimation techniques were developed to mitigate
the effects of urban multipath and signal blockage, and to recover from false
integer fixes. The estimator was evaluated using the publicly-available TEX-CUP
urban positioning dataset, yielding a \baselinelynxPvShort{} and
\baselinelordPvShort{} integer fix availability, and
\baselinelynxpnfHorzTotal{} cm and \baselinelordpnfHorzTotal{} cm overall (fix
and float) 95-th percentile horizontal positioning error with a consumer-grade
and industrial-grade inertial sensor, respectively, over more than two hours of
driving in the urban core of Austin, Texas. A performance sensitivity analysis
showed that the false-fix detection and recovery scheme is key to achieving an
acceptably low false integer fixing rate of \baselinelordPfShort and
\baselinelynxPfShort, respectively. Having a second vehicle-mounted GNSS
antenna significantly increased integer-fix availability, decreased false-fix
rate, and improved both root-mean-square and 95th-percentile positioning
performance as compared to a single-baseline CDGNSS configuration.

\section*{Acknowledgments}
This work was supported by the U.S. Department of Transportation under Grant
69A3552047138 for the CARMEN University Transportation Center, and by the Army
Research Office under Cooperative Agreement W911NF-19-2-0333. The views and
conclusions contained in this document are those of the authors and should not
be interpreted as representing the official policies, either expressed or
implied, of the Army Research Office or the U.S. Government. The
U.S. Government is authorized to reproduce and distribute reprints for
Government purposes notwithstanding any copyright notation herein. Map data
\ifthesis\else{\textcopyright{}\;}\fi  OpenStreetMap contributors
(\url{https://www.openstreetmap.org/copyright}).

\bibliographystyle{ieeetr} 
\bibliography{pangea}
\end{document}